\documentclass[twocolumn]{aastex63}

\usepackage{color}
\usepackage{gensymb}
\usepackage{enumitem}

\newcommand\swift{{\it Swift}}

\newcommand\xmm{{\it XMM-Newton}}

\newcommand\astrosat{{\it AstroSat}}
\newcommand\nustar{{\it NuSTAR}}
\newcommand\integral{{\it INTEGRAL}}
\newcommand\hst{{\it HST}}

\newcommand\s{{\rm~s}}
\newcommand\ks{{\rm~ks}}

\newcommand\mpc{{\rm~Mpc}}

\newcommand\kev{{\rm~keV}}
\newcommand\ev{{\rm~eV}}
\newcommand\ergs{{\rm~ergs}}
\newcommand\cm{{\rm~cm}}
\newcommand\km{{\rm~km}}
\newcommand\angstrom{{\rm~\AA}}

\shortauthors{Tripathi \& Dewangan}

\begin{document}


\title{\astrosat{} view of spectral transition in the changing-look active galaxy NGC~1566 during the declining phase of the 2018 outburst}

\correspondingauthor{Prakash Tripathi}
\email{prakasht@iucaa.in}


\author[0000-0003-4659-7984]{Prakash Tripathi}
\affiliation{Inter University Centre for Astronomy and Astrophysics, Pune, India, 411007}

\author[0000-0003-1589-2075]{Gulab C. Dewangan}
\affiliation{Inter University Centre for Astronomy and Astrophysics, Pune, India, 411007}

\begin{abstract}
 
 NGC~1566 is a changing-look active galaxy that exhibited an outburst during 2017--2018 with a peak in June 2018. We triggered \astrosat{} observations of NGC~1566 twice in August and October 2018  during its declining phase. Using the \astrosat{} observations along-with two \xmm{} observations in 2015 (pre-outburst) and June 2018 (peak-outburst), we found that the X-ray power-law, the soft X-ray excess, and the disk components showed extreme variability during the outburst. Especially, the soft excess was negligible in 2015 before the outburst, it increased to a maximum level by a factor of $>200$ in June 2018, and reduced dramatically by a factor of $\sim7.4$ in August 2018 and become undetectable in October 2018. The Eddington fraction ($L/L_{Edd})$ increased from $\sim0.1\%$ (2015) to $\sim 5\%$ (June 2018), then decreased to $\sim 1.5\%$ (August 2018) and $0.3\%$ (October 2018). Thus, NGC~1566 made a spectral transition from a high soft-excess state to a negligible soft-excess state at a few $\%$ of the Eddington rate. The soft-excess is consistent with warm Comptonization in the inner disk that appears to have developed during the outburst and disappeared towards the end of the outburst over a timescale comparable to the sound crossing time. The multi-wavelength spectral evolution of NGC~1566 during the outburst is most likely caused by the radiation pressure instability in the inner regions of the accretion disk in NGC~1566.

\end{abstract}


\section{Introduction} 
\label{sec_intro}

Active Galactic Nuclei (AGN) are luminous objects in the Universe that vary across the entire electromagnetic band from radio to X-rays and $\gamma$-rays. A small number of AGN exhibit large amplitude variability or outbursts in their optical/ultraviolet (UV) and X-ray luminosity  on timescales of months to a few years. These sources change their spectroscopic appearance due to strongly variable broad Balmer emission lines, and are classified as "changing-look" AGN (CL-AGN, \citealt{2003MNRAS.342..422M}). Observationally, the number of CL-AGN is growing \citep{2016ApJ...826..188R, 2018MNRAS.480.4468R, Yang_2018, 2018ApJ...864...27S, 2019ApJ...874....8M}. The drastic variations in the optical/UV continuum, broad emission lines, and the X-ray spectrum of CL-AGN result in a change in the formal classification of their Seyfert types \citep{2005A&A...442..185B, 2014ApJ...788...48S}. The physical mechanism responsible for the multi-wavelength variability in CL-AGN is not well understood. The possible explanations are, radiation pressure instabilities in the disk \citep{2020A&A...641A.167S}, changing the inner disk radius leading to state transition \citep{2018MNRAS.480.3898N, 2019ApJ...883...76R}, tidal disruptive events (TDEs) \citep{2020ApJ...898L...1R}, variation in the accretion rate \citep{2014MNRAS.438.3340E}, and variable obscuration causing a switch from a Compton-thick to Compton-thin absorption in the X-ray band \citep{2002MNRAS.329L..13G, 2003MNRAS.342..422M}. 
For most of the sources, the changing-look behavior has been attributed intrinsic to the source \citep{2019A&A...625A..54H, 2018ApJ...866..123M, 2017ApJ...846L...7S, 2018ApJ...864...27S}.  

NGC~1566 is a nearby ($\rm z=0.00502$)
Seyfert galaxy and the nearest CL-AGN that consists of a super-massive black-hole (SMBH) at the center with the black-hole mass estimated using various techniques of $\log(M_{BH}/M_{\odot})\sim 6.92$ (stellar velocity dispersion; \citet{2002ApJ...579..530W}), $7.11\pm0.32$ (spiral pitch angle; \citet{2014ApJ...789..124D}), $6.83\pm0.3$ (molecular gas dynamics; \citet{2019A&A...623A..79C}), {and  $6.48^{+0.11}_{-0.15}$  
(Br$\gamma$ emission line; \citet{2015A&A...583A.104S})}.
This source exhibits recurrent outbursts. Based on spectrophotometric monitoring of NGC~1566 nucleus during 1970--1985, \cite{1986ApJ...308...23A} reported four periods of the outburst activities over the 15 years long baseline. Each of them lasted for $\sim 1300$ days with a quiescent phase of a few months to years between two consecutive burst periods. The AGN remained at the highest flux state for less than a month. In the activity period 1981 January to 1985 August,  \cite{1986ApJ...308...23A} found a series of rapid bursts, each with a typical rise time of $\sim 20$ days and a long exponential decay of $\sim 400$ days.  \citet{2019MNRAS.483L..88P} noticed another single outburst  in the 105-months long \swift{}/BAT\footnote{\url{https://swift.gsfc.nasa.gov/results/bs105mon/216}} light-curve covering the period from December 2004 to August 2013. During this outburst in 2010, the source attained an outburst peak of 4~mCrab from a low flux level of 0.8~mCrab in the $15-200\kev$ band in $2-3$ months. Further, in 2018 an increase in the hard X-ray flux was observed by \integral{}~\citep{2018ATel11754....1D} and followed by \swift{}~\citep{2018ATel11783....1F}. \cite{2019MNRAS.483..558O} mention that the beginning of this burst is uncertain due to the lack of data during the rising period but the light-curves in {\it ASAS-SN} V-band and {\it NEOWISE} mid-infrared showed that the flux started increasing around September 2017 \citep{2018ATel11893....1D,2018ATel11913....1C}.  
After a  rising period of $\sim9$ months the source reached to the peak of the burst around {June--July} 2018 \citep{2019MNRAS.483L..88P,2019MNRAS.483..558O,2020MNRAS.498..718O}. 

\cite{2019MNRAS.483..558O, 2020MNRAS.498..718O} analyzed the UV and X-ray data of NGC~1566 acquired with \swift{} and the MASTER Global Robotic Network from 2007 to 2019. During this period, the source was in a low flux state till 2015 followed by an observational gap during 2015 to 2018. The flux of the source started rising and a strong outburst was observed in 2018, peaking during the end of June -- beginning of July 2018. A substantial increase in the X-ray flux by a $\sim 1.5$ order of magnitude was observed followed by an increase in the UV flux. The maximum measured X-ray flux of $1.1\times10^{-10}\ergs\cm^{-2}\s^{-1}$ on MJD: 58309 is $\sim 50$ times larger than the minimum recorded flux of $2.2\times10^{-12}\ergs\cm^{-2}\s^{-1}$ on MJD: 56936 \citep{2019MNRAS.483..558O}. Using the post-brightening observations, \cite{2020MNRAS.498..718O} reported 3 more re-brightening states of the source during the decline of the outburst on ($i$) November 17, 2018 to January 10, 2019 (MJD : 58440--58494), ($ii$) April 29, 2019 to June 19, 2019 (MJD : 58603--58654), and ($iii$) July 27, 2019 to August 6, 2019. They found the soft X-ray band ($0.5-2\kev$) to be more variable (by a factor of $\sim 30$)  than the hard X-ray  ($2-10\kev$) band (by a factor of $\sim20$) during the 9 months after the maximum of the burst. \cite{2019MNRAS.483L..88P} investigated the X-ray spectral properties of the source using simultaneous \xmm{} and \nustar{} observations performed on 26 June 2018 when the outburst was at the peak, and \xmm{} data alone acquired on 6 November 2015 when the source was at a low flux state. The July 2018 X-ray spectrum showed a power-law component, soft X-ray excess emission, distant and relativistic reflection components, and two warm absorbing components with low  column densities ($N_H\sim 2.5\times10^{20}\cm^{-2}$). The quiescent X-ray spectrum of 2015 observation is dominated by AGN with negligible contribution from extended emission. They estimated the Eddington ratios of $0.2\%$ and $\sim 5\%$ for the 2015 and 2018 data using the 2--10\kev{} X-ray flux, a bolometric correction factor of 20, { and the black-hole mass from \cite{2002ApJ...579..530W}. They found the black-hole spin parameter to be $a< 0.25$. }

Here, we investigate the FUV/X-ray spectral evolution of the source during the decline period of the {June--July} 2018 outburst using observations performed with \astrosat{}~\citep{2014SPIE.9144E..1SS}. We also analyze the UV/X-ray \xmm{} data from the 2015 and 2018 observations to find the evolution of different spectral components when compared with our \astrosat{} observations. While the \xmm{} X-ray data are already analyzed by \citep{2019MNRAS.483L..88P}, the joint UV/X-ray broadband study has not been performed using these observations. 
We organise the paper as follows. We describe the observation and data reduction methods in section~\ref{sec_obs}. We describe the spectral analyses in section~\ref{sec_analysis} followed by results and discussion in section~\ref{sec_results}. Finally, we conclude our results in section~\ref{sec_conclusion}.
 
\section{Observation and Data reduction}
\label{sec_obs}
We triggered Target of Opportunity (ToO) observations of NGC~1566 with the \astrosat{}~\citep{2014SPIE.9144E..1SS} mission during the declining phase of 2018 outburst.
We observed the source on August 11, 2018 (MJD : 58340.46, ObsID: T02\_085T01\_9000002296, hereafter obs1) and on October 22, 2018 (MJD : 58412.78, ObsID: T03\_020T01\_9000002444, obs2). These observations were performed after $\sim 30$ days of the maximum of the 2018 burst (MJD: 58309) and few days before the post-brightening bursts reported by \cite{2020MNRAS.498..718O}. \astrosat{} carries 4 co-aligned payloads -- the Large Area X-ray Proportional Counter (LAXPC; \citealt{2016SPIE.9905E..1DY, 2017JApA...38...30A, Antia_2017}), the Soft X-ray Telescope (SXT; \citealt{10.1117/12.2235309, 2017JApA...38...29S}), the Ultra-Violet Imaging Telescope (UVIT; \citealt{Tandon_2017, 2020AJ....159..158T}), and the Cadmium-Zinc-Telluride Imager (CZTI; \citealt{2016SPIE.9905E..1GV}) that observe simultaneously. We used the data observed with the SXT and UVIT. We did not use the the LAXPC data due to the high background and the CZTI  data as the source was not detected. Details of the observations are listed in Table~\ref{tabobs}.

\subsection{The SXT data}

The SXT is a focusing X-ray telescope with a CCD camera operating in the photon counting mode. It is capable of low resolution imaging (FWHM $\sim 2{\arcmin}$, HPD $\sim 11{\arcmin}$) and medium resolution spectroscopy (FWHM $\sim 150\ev$ at $6\kev$) in the $0.3-8\kev$ band. We processed the level-1 data using the SXT {data processing} software (AS1SXTLevel2-1.4b) available at the SXT payload operation center (POC) website\footnote{\url{https://www.tifr.res.in/~astrosat_sxt/sxtpipeline.html}}, and generated level-2 clean event files for individual orbits of each observation. We merged the clean event files for a given observation id using the {\it Julia} SXT event merger tool ({\textsc{sxt\_l2evtlist\_merge}}) developed by us and made available at the SXT POC website. We obtained the processed and cleaned level-2 data for a net exposure time of $\sim 23.3\ks{}$ (obs1) and $\sim 19.1\ks{}$ (obs2).  We used \textsc{xselect} tool  and extracted the source spectrum from the two merged level-2 event files using a circular region of $15\arcmin$ radius centered on the source position. 
The large HPD ($\sim 11\arcmin$) of the  SXT and the calibration sources present at the corners of the CCD camera do not leave source-free regions. Therefore, we used the blank sky spectrum ({SkyBkg\_comb\_EL3p5\_Cl\_Rd16p0\_v01.pha}) as the background spectrum as recommended by the SXT POC. We also used the recommended redistribution matrix file ({sxt\_pc\_mat\_g0to12.rmf}) and the ancillary response file ({sxt\_pc\_excl00\_v04\_20190608.arf}) for the SXT spectral analysis. We grouped the SXT spectral data to a minimum of 25 counts per bin using the {\textsc{grppha}} tool. The background-corrected net source count rates are $0.637\pm0.006$ and $0.186\pm0.004$ counts~s$^{-1}$ for obs1 and obs2, respectively. Thus, the observed X-ray count rate  decreased by a factor of $\sim 3.4$ in $\sim 70$ days.


\begin{figure*}
	\centering
	\includegraphics[scale=0.38]{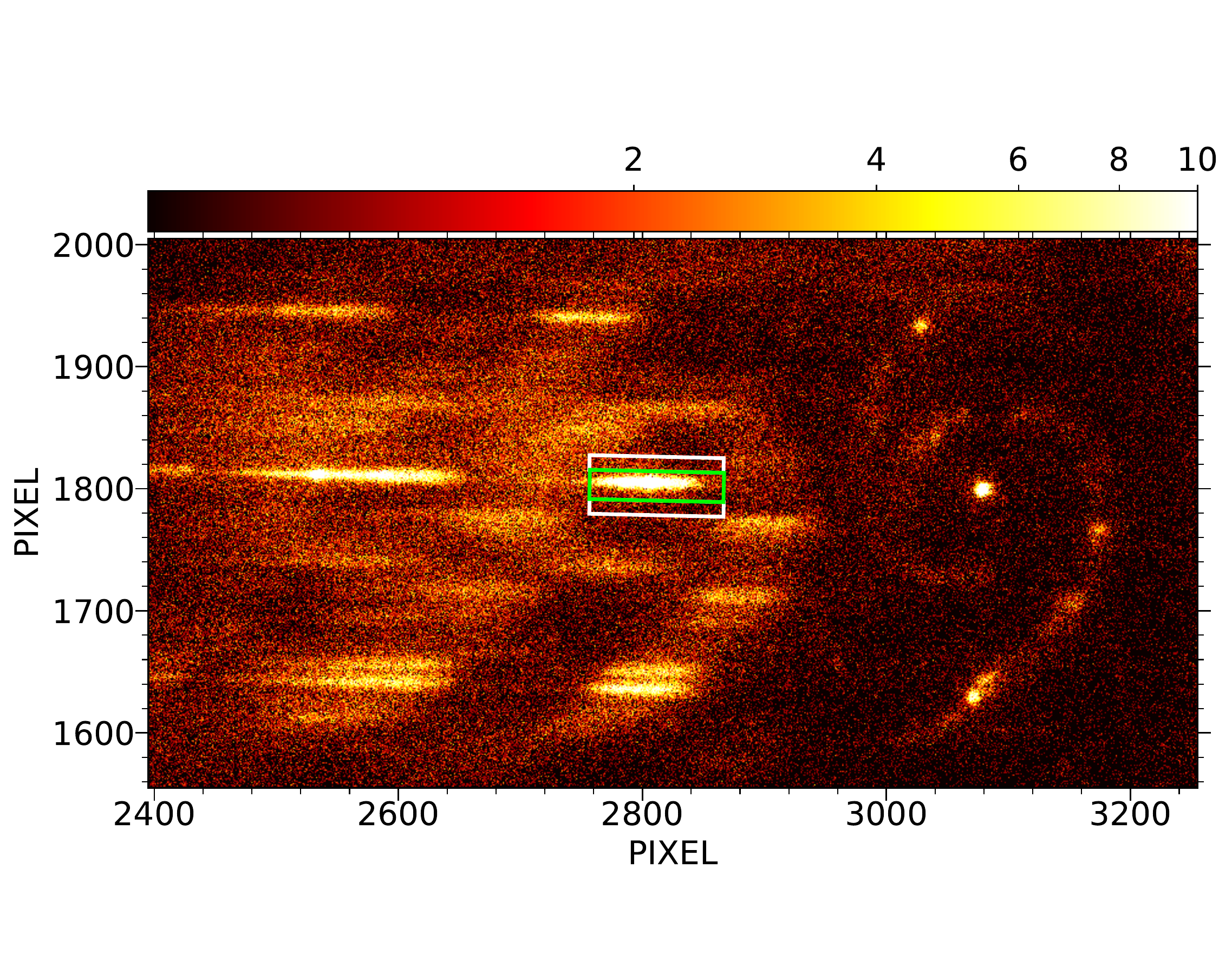}
	\includegraphics[scale=0.38]{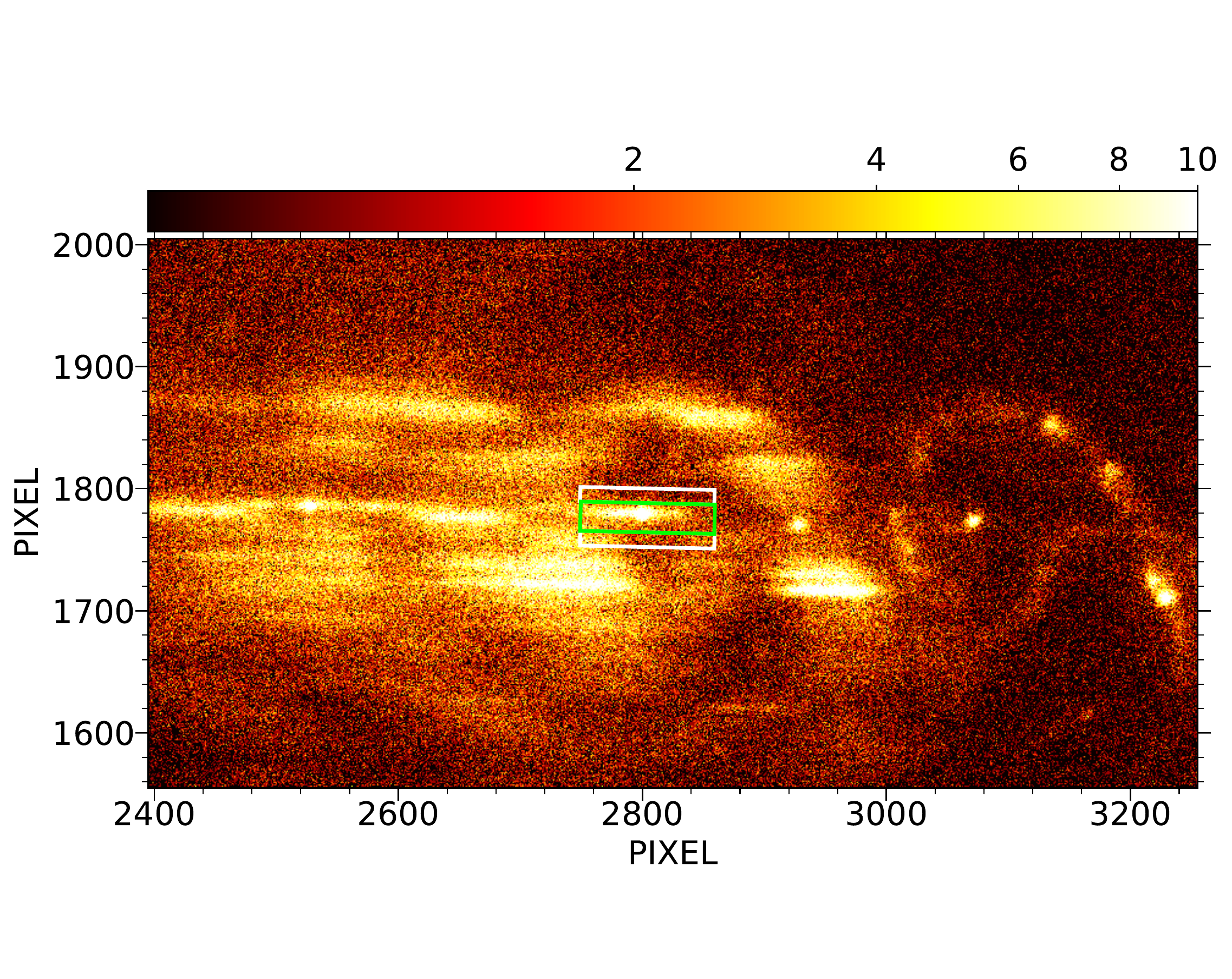}
	\caption{The \astrosat{}/UVIT FUV-G1 images from obs1 (ObsID T02\_085T01\_9000002296) and obs2 (ObsID T03\_020T01\_9000002444) in the {\it left} and  {\it right} panels, respectively. The first and second orders are relatively clean in obs1 than those of obs2. In obs2, the second order is strongly contaminated, while the first order is relatively free of contamination. {The green and white rectangles represent the source and background extraction regions, respectively. The images are shown at the same color scale, and the color-bars are shown at the top of the images. }}
	\label{figgratim}
\end{figure*}
 
\begin{table*}
	\centering
	
		\caption{\astrosat{} observations of NGC~1566. The net source and background count rates with $1-\sigma$ errors are listed.}
		\label{tabobs}
		
		\begin{tabular}{ccccccccc}
			\hline\hline 
		Date	& Obs. ID	&\multicolumn{5}{c}{UVIT}	&\multicolumn{2}{c}{SXT}\\
		    &&\multicolumn{2}{C}{\rm~{FUV-G1}}&\multicolumn{3}{c}{F154W}&&\\
\hline
	dd/mm/yyyy&		&Exp	&Rate$^a$	&Exp	&Rate$^b$ &Bkg$^c$ &Exp	&Rate$^d$ \\	
				&		&(ks) &(cts~s$^{-1}$)	&(ks) &(cts~s$^{-1}$)	& (cts~s$^{-1}$) &(ks) &(cts~s$^{-1}$)\\

			\hline
	11/08/2018	& T02\_085T01\_9000002296	&2.5	&$4.69\pm0.05$ 	&3.5 &$11.03\pm0.07$ &$0.27\pm0.04$ &23.3	&$0.637\pm0.006$ \\
	22/10/2018	& T03\_020T01\_9000002444	&4.8	&$1.21\pm0.03$ 	&1.5    &$3.56\pm0.06$ &$0.23\pm0.06$  &19.1	&$0.186\pm0.004$ \\
			\hline\hline

		\end{tabular}
		\\
$^a$ = Background-subtracted net source count rate derived from the 1st order of FUV-G1 in the 1225--1852\AA band.\\
$^b$ = The mean background-subtracted net source count rate of F154W filter extracted from a circular region of $7\arcsec$ radius (shown in green circle in Figure~\ref{figfilter}).\\
$^c$ = The mean background count rate derived from 3 circular regions of $7\arcsec$ radius each (shown in white circles in Figure~\ref{figfilter})\\
$^d$ = Background-subtracted net source count rate in 0.5--6.5 keV band extracted from a circular region of 15$\arcmin$ radius \\
\end{table*}

\subsection{The UVIT data}

The UVIT \citep{Tandon_2017, 2020AJ....159..158T} consists of two co-aligned telescopes, one of them  is sensitive in the far {ultra-violet} (1300--1800\angstrom{}) band and is referred as the FUV channel.  The light from the second telescope is split into {the near ultra-violet} (2000--3000\angstrom{}) and visible  (3200--5500\angstrom{}) bands, thus forming the NUV and VIS channels, respectively. The FUV and NUV channels are equipped with a number of broadband filters and operate  in the photon counting mode, and provide high resolution (FWHM = $1-1.5\arcsec$) imaging capability. Additionally, the two FUV gratings and one NUV grating mounted in the filter wheels are useful for low resolution slit-less spectroscopy. 
The maximum efficiency is achieved in the $-2$ order of the FUV gratings and $-1$ order of the NUV grating. The peak effective area and the spectral resolution (FWHM) are $\sim 4.5 \cm^{-2}$ and $\sim 15$\angstrom{} for the FUV gratings and $\sim 18.7\cm^{-2}$ and $\sim 33$\angstrom{} for the NUV grating \citep{2021JApA...42...49D}. We observed NGC~1566  using both the FUV gratings in obs1 and only the FUV-Grating1 (FUV-G1) in obs2. We could not observe the source in the NUV band as the NUV channel stopped functioning in March 2018 \citep{2021JApA...42...20G}. We processed the level-1 FUV data using the UVIT pipeline {\textsc{ccdlab}}~\citep{2017PASP..129k5002P} software, and generated clean merged images for each grating/filter used in  obs1 and obs2. We use the FUV-G1 data here as the FUV-G2 data from obs1 is severely contaminated by other sources along the dispersion direction. The observational details are listed in Table~\ref{tabobs}. 

In Figure~\ref{figgratim}, we show the FUV-G1 images of obs1 and obs2 in the left and right panels, respectively. The two observations were performed at different roll angles. Therefore, the orientation of the galaxy in the detector plane is changed such that the spiral arms are rotated by $\sim80\degree$ anticlockwise {in obs2 relative to the obs1}. A star present in the upper spiral arm in obs1 is rotated in obs2 that resulted in a stronger host galaxy contamination in obs2. In the right panel of Figure~\ref{figgratim}, $-2$ order spectrum  of NGC~1566 nucleus  is strongly contaminated by the $-1$ and $-2$ orders of a strong UV source located in one of the spiral arms, whereas $-1$ order of NGC~1566 nucleus is relatively less contaminated (see Figure~\ref{figgratim}). Therefore, we extracted the source spectrum from $-1$ order of FUV-G1 with a cross-dispersion width of 24 pixels (where, 1 pixels $\sim 0.41\arcsec$), and the background spectra from above and below the AGN region with a cross-dispersion width of 12 pixels for each region. We selected the background regions exactly in the same pixel range along the dispersion direction as for the AGN region, but above and below the dispersed image of the AGN. {The source and background extraction regions are shown as the green and white rectangles, respectively, in Figure~\ref{figgratim}}. We used the UVITTools package \citep{2021JApA...42...49D} written in the {\it Julia} language to extract the 
{ OGIP compliant} spectral files for a given grating order and cross-dispersion width \citep{2021JApA...42...49D}. We combined the two background spectra for a single observation using {\textsc{addspec}}\footnote{https://heasarc.gsfc.nasa.gov/ftools/caldb/help/addspec.txt} task. We used the instrument response file ({fuv\_grating1\_m1\_3oct19.rmf}; \citealt{2021JApA...42...49D}) in the spectral analysis. {We did not require to group the data as each energy bin of the spectrum has more than $20$ counts}. We list the net source count rates for FUV-G1 in Table~\ref{tabobs}. The FUV-G1 count rate varied by a factor of $\sim 3.9$. 
 
\begin{figure*}
    \centering
    \includegraphics[scale=0.38]{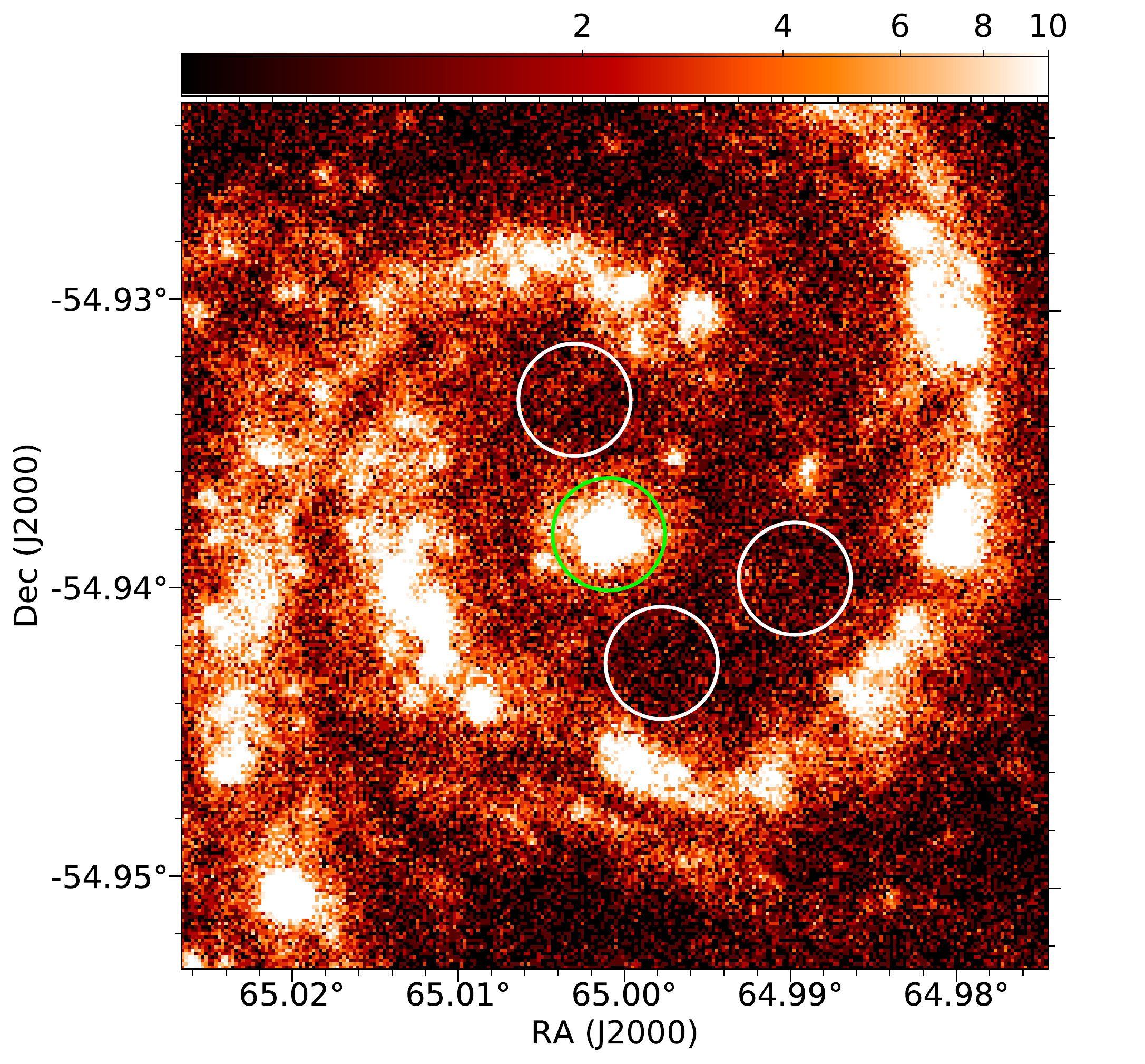}
    \includegraphics[scale=0.38]{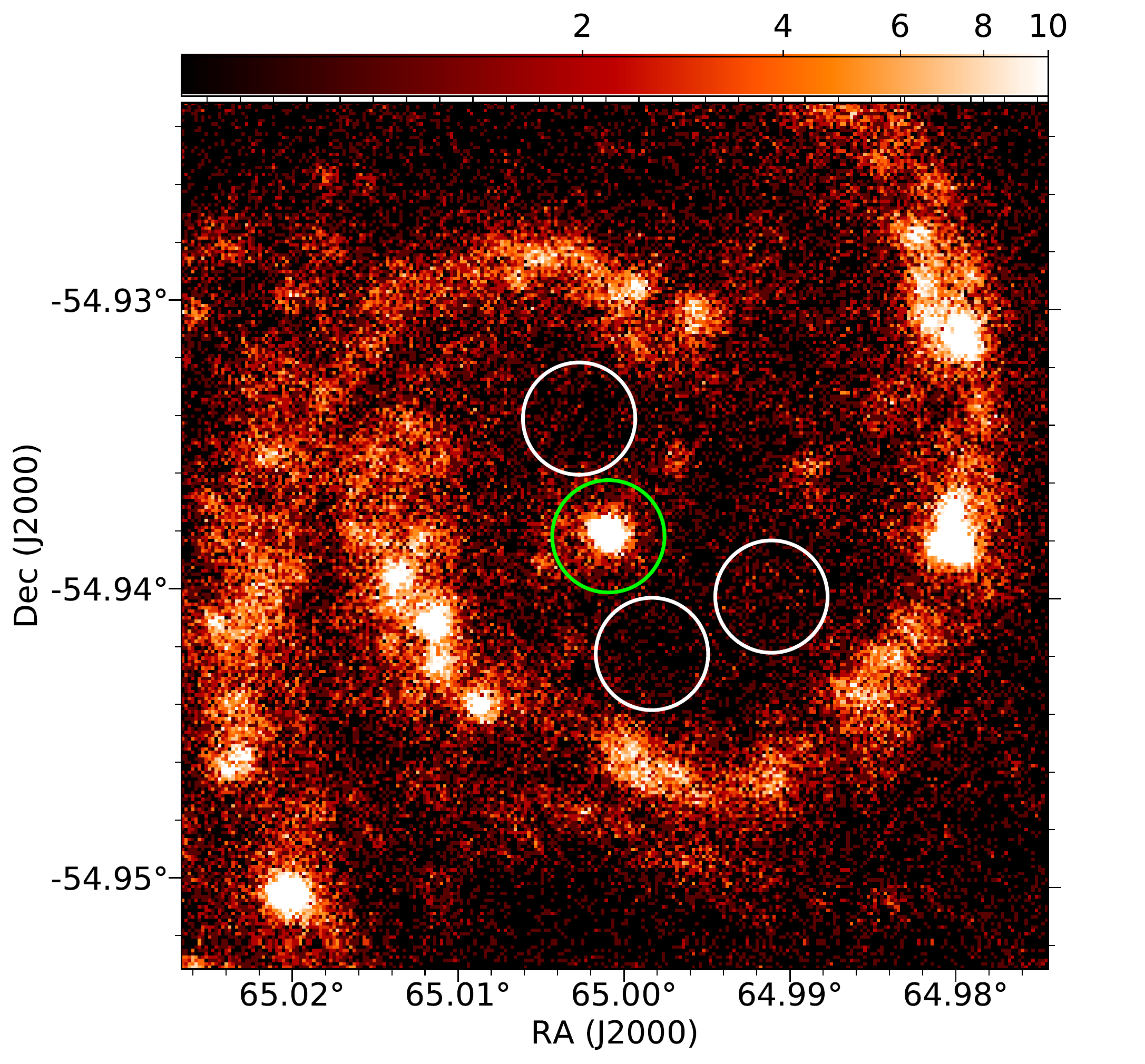}
    \caption{Source (green circles) and background (white circles) regions are shown in the F154W images of obs1 (left) and obs2 (right). Each circle has a radius of 7\arcsec. {The images are shown at the same color scale, and the colorbars are shown at top of the images.} }
    \label{figfilter}
\end{figure*}
 
 
 Figure~\ref{figfilter} shows the broadband images of NGC~1566 in the F154W filter (mean wavelength  $\lambda_m = 1541${\rm~\AA}, $\Delta\lambda = 380${\rm~\AA}).  The central AGN is clearly separated from the nearby sources as well as from the spiral arms of the galaxy resulting in a relatively less host galaxy contamination compared to the FUV-Grating data. Therefore, we also used the broadband filter data to derive the relatively uncontaminated AGN emission which we use to scale  the slit-less grating spectra. We extracted the source counts from a circular region of $7\arcsec$ radius centered at the location of the nucleus (shown as green circles in Figure~\ref{figfilter}). The excellent spatial resolution of the UVIT allows us to choose such a small extraction region that includes almost all AGN emission but avoids significant contamination from the host galaxy emission. To correct for the background as well as the host galaxy emission, we extracted background counts from three local source-free, circular regions with $7\arcsec$ radii (shown as white circles in Figure~\ref{figfilter}) and calculated the mean background level for each observation. We subtracted the mean background rate from the source count rate, and derived the net source count rate for both observations. We have listed the net source count rates and the mean background rates in Table~\ref{tabobs}. The net source count rate decreases from $11.03\pm0.07$ to $3.56\pm0.06$ counts~s$^{-1}$ in $\sim 70$ days. We wrote the net source count rate in a {\textsc{xspec}} compatible spectral file using {\textsc{ftflx2xsp}} task for each observations and used the F154W filter response ({F154W\_effarea\_Tandon\_etal2020.rsp}; \citealt{2021JApA...42...49D}) in our spectral analysis below.


\section{Spectral Analysis}
\label{sec_analysis}
We performed spectral analysis using {\textsc{xspec}} \citep{1996ASPC..101...17A} and {\textsc{sherpa}} \citep{2001SPIE.4477...76F} packages. We used the Galactic column density of $N_H = 9.19\times 10^{19}\cm^{-2}$ \citep{2005A&A...440..775K}, and the abundance and the cross-section of inter stellar medium according to \cite{2009ARA&A..47..481A} and \cite{1996ApJ...465..487V}, respectively. We used the $\chi^2$ - minimization technique to obtain the best-fitting model. We quote  the errors on the best-fit parameters at the $90\%$ confidence level.


\subsection{SXT spectral analysis}

\begin{figure}
	\centering
	\includegraphics[scale=0.42]{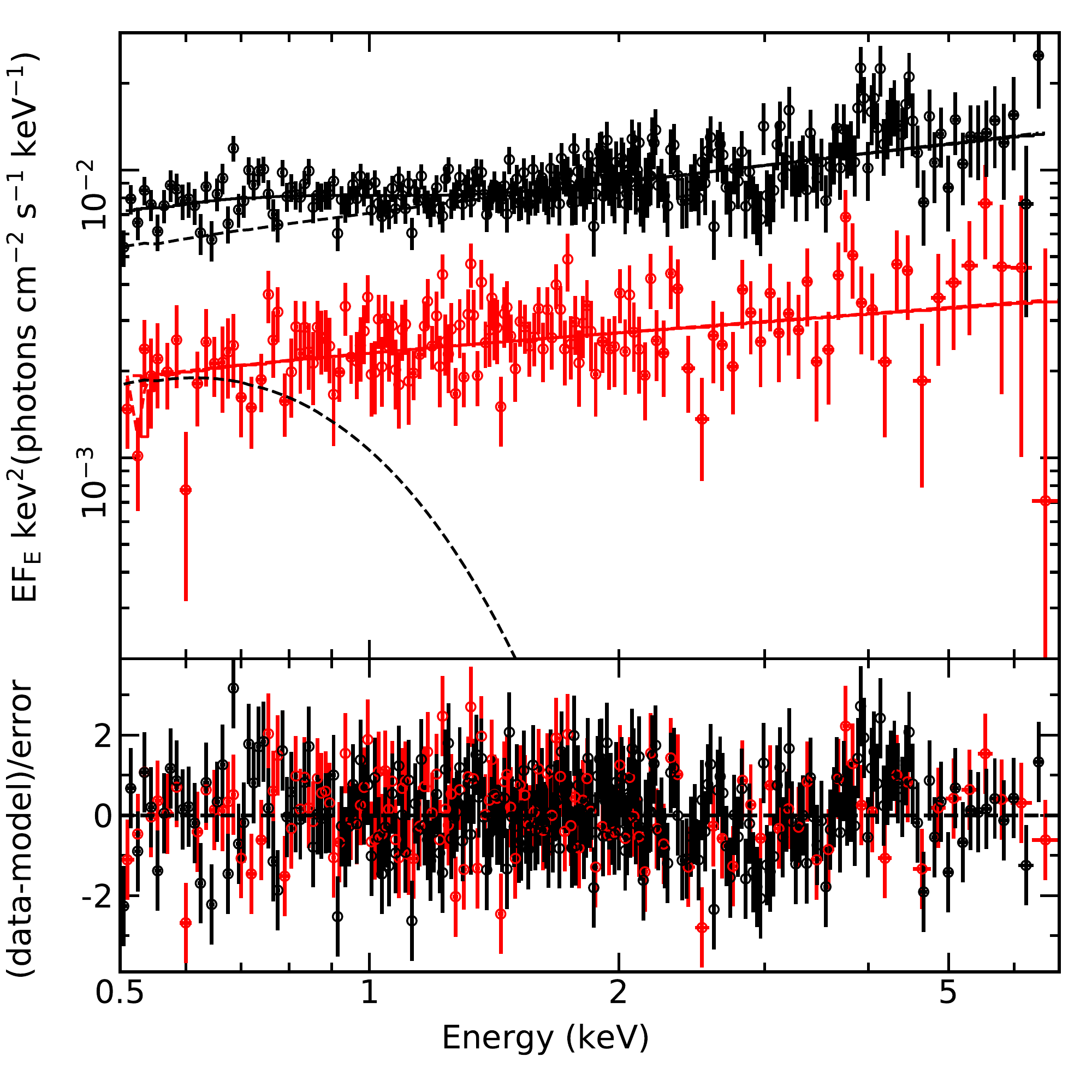}
	\caption{The best-fit SXT (0.3--6.8\kev{}) spectra of obs1 (black) and obs2 (red) fitted with {\textsc{tbabs$\times$(zpowerlaw+zbbody)}} models. The total models (solid), {\textsc{zpowerlaw}} (dashed), and {\textsc{zbbody}} (dashed) models for a single observation are shown with the same colors. The {\textsc{zbbody}} model is negligible for obs2. The bottom panel show residuals in terms of (data-model)/error.}
	\label{figsxtfit}
\end{figure}

We begin our spectral analysis by fitting the two SXT data sets jointly in the $0.5-6.8\kev$ band. 
After the launch of \astrosat{}, gain of the SXT CCD has drifted slightly which we handle using the {\textsc{gain}} command in {\textsc{xspec}}. We fixed the slope at 1.0 and varied the intercept. We used an absorbed power-law model ({\textsc{tbabs$\times$zpowerlaw}} in {\textsc{xspec}} terminology) to fit the data. We applied 3\% systematic error to account for uncertainties in the calibration and background. We varied photon-index and normalization of the {\textsc{zpowerlaw}} model across the 2 datasets. The fit resulted in $\chi^2=439.3$ for 400 degrees of freedom ($dof$). We noticed the soft X-ray excess emission below 2\kev{} in the best-fit residuals. Therefore, we added a black-body component {\textsc{zbbody}} to the model. Initially, we tied the temperature and the normalization of the {\textsc{zbbody}} between the two data sets. This resulted in a poor-fit with $\chi^2/dof=433.9/398$. We then varied the normalization of the {\textsc{zbbody}} component for the two data sets, independently. The fit improved significantly with $\Delta{\chi^2} = -16.1$ for 1 parameter ($\chi^2/dof=417.8/397$). An F-test in {\textsc{xspec}}  resulted in the F-statistic of 15.3 and probability of $\sim10^{-4}$ suggesting that a variation in the black-body normalization is statistically significant. The black-body normalization for the obs2 is not well constrained and is consistent with zero, implying the absence of the soft X-ray excess in this observation. We also varied the black-body temperature between the two datasets that did not improve the fit ($\Delta{\chi^2}=0.5$ for 1 parameter). The best-fit photon-index and the $2-10\kev$ flux of the X-ray power-law component are $\Gamma = 1.65^{+0.06}_{-0.08}$ and $f_{PL}=3.2\pm0.2\times10^{-11}\ergs\cm^{-2}\s^{-1}$ for the obs1, and $\Gamma=1.78^{+0.07}_{-0.13}$ and $f_{PL}=0.85\pm0.08\times10^{-11}\ergs\cm^{-2}\s^{-1}$ for the obs2. The black-body temperature and $0.5-2\kev$ flux are $kT_{bb}=0.16\pm0.02\kev$, and $f_{BB} = 2.56\pm1.03\times10^{-12}\ergs\cm^{-2}\s^{-1}$ for the obs1, and $0.45\times10^{-12}\ergs\cm^{-2}\s^{-1}$ ($90\%$ upper limit) for the obs2. The soft X-ray excess and the power-law flux varied by factors of { $>5.7$ and $\sim3.8$} in $\sim 70$ days. The SXT spectral data, the best-fit models and the deviations of the  data from the models are shown in Figure~\ref{figsxtfit}. We also calculated the observed flux in the $0.5-10\kev$ band to be $4.9\times10^{-11}$ and $1.4\times10^{-11}\ergs{}\cm^{-2}\s^{-1}$ for the obs1 and obs2, respectively, which are consistent for the intermediate states between the maximum and minimum of the outburst \citep{2019MNRAS.483..558O}.

The soft X-ray excess can be produced from thermal Comptonization of the disk seed photons in a warm and optically thick corona. We tested this scenario by modeling the soft X-ray excess emission with a thermal Comptonization model, {\textsc{nthcomp}} \citep{nthcomp1, nthcomp2}. The main parameters of this model are photon-index of the Comptonized emission ($\Gamma_{warm}$), temperature of the warm corona ($kT_{warm}$), and temperature of the seed photons ($kT_{seed}$). We assume the black-body seed photons (inp\_type = 0) with a temperature of ($kT_{seed}=1\ev$). We used this model for the first epoch only where the soft X-ray excess is present. The model {\textsc{tbabs$\times$(zpowerlaw+nthcomp)}}  resulted in a similar fit quality ($\chi^2/dof=417.7/397$) as before. The best-fit $\Gamma_{warm}$, $kT_{warm}$, and normalization ($N_{nthcomp}$) of the {\textsc{nthcomp}} model are $\Gamma_{warm}\leq 3$ ($3\sigma$ upper limit), $kT_{warm} = 0.15^{+0.17}_{-0.02}\kev$, and $N_{nthcomp} = 1.2^{+1.3}_{-0.6}\times10^{-3}$. The soft X-ray excess flux in the $0.5-2\kev$ band using the best-fit {\textsc{nthcomp}} component is $f_{SXE}=3.4\pm 1.8\times10^{-12}\ergs\cm^{-2}\s^{-1}$. 

\subsection{UVIT/FUV grating spectral analysis}

\begin{figure}
	\centering
	\includegraphics[scale=0.42]{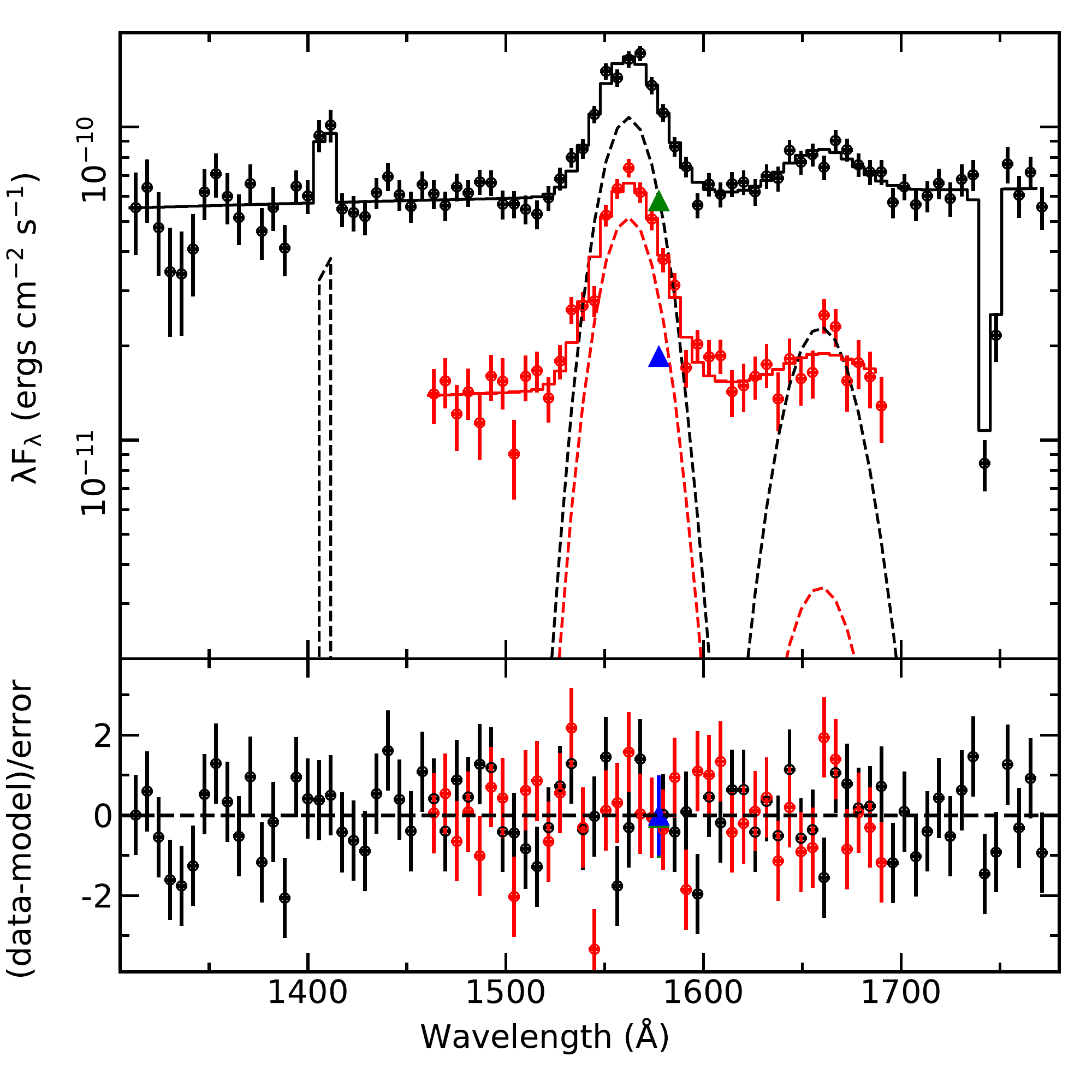}
\caption{FUV spectra of NGC~1566 derived using FUV-G1 and F154W data of 2 observations fitted with {\textsc{const$\times$redden$\times$gabs$\times$(zpowerlaw+zgauss+zgauss+ zgauss)}} model. The FUV-G1 and F154W data are plotted in black and green colors for obs1, and in red and blue colors for obs2, respectively. The bottom panel shows $\chi$ residual in terms of (data-model)/error.}
	\label{figfuvfit}
\end{figure}

We analyzed the two sets of FUV-G1 grating + F154W filter data from the two \astrosat{} observations jointly in {\textsc{xspec}}. Inspecting the grating data, we noticed strong contamination from the diffuse emission of the host galaxy at the low and high ends of the FUV band ($1250-1800$\angstrom{}). The contamination is stronger for obs2. We could use the data in the $1305-1771$\angstrom{} ($7-9.5\ev$) range for obs1, and in the $1458-1698$\angstrom{} ($7.3-0.85\ev$) range for obs2. 
The broadband F154W data are relatively clean from the host galaxy contamination.  Therefore, we analyzed the grating and the broadband data jointly, and  our UV flux measurements refer to the F154W data.

We used a simple {\textsc{zpowerlaw}} model modified by the Galactic reddening with {\textsc{redden}}~\citep{1989ApJ...345..245C} component. 
We fixed the color excess parameter of the {\textsc{redden}} component at $E(B-V)=0.0134$ calculated from the Galactic column density of $\rm N_H = 9.19\times10^{19}\cm^{-2}$ using the gas-to-dust ratio $N_H = E(B-V)\times6.86\times 10^{21}$ \citep{2009MNRAS.400.2050G}. We also used a {\textsc{constant}} component to account for relative normalization between the grating and the filter data. We fixed the {\textsc{constant}} at 1 for the filter and varied it for the grating for each observation. This relative normalization is useful to account for any differences in the host galaxy contamination and/or differences in the absolute flux calibration \citep{2021JApA...42...49D}.  We kept tied the rest of the model parameters across the FUV-G1 and F154W data for each observation. We varied the photon-index and the normalization of the {\textsc{zpowerlaw}} model for both observations. This resulted in a poor-fit with $\chi^2/dof=1125.2/116$. We noticed strong residuals near $1550$\angstrom{} due to \ion{C}{4}$\lambda1549$ emission line. We added a Gaussian line ({\textsc{zgauss}}) to the model. We tied the energy and the width ($\sigma$) of the line across the two data sets and varied the normalization for each observation. The fit improved significantly with $\chi^2/dof=185.9/112$. We further noticed an absorption feature near 1741\angstrom{} (7.15\ev{}) { most likely associated with \ion{Ni}{2} \citep{1999MNRAS.310..289O}}. Therefore, we added a Gaussian absorption line ({\textsc{gabs}}). The fit further improved to $\chi^2/dof=134/109$. We also noticed a weak residual near 1402\angstrom{} (8.9\ev{}) most likely due to \ion{Si}{4} and \ion{O}{4}] emission lines. We noticed this line in obs1 only as the data below $1450\angstrom{}$ were excluded in obs2 due to the host galaxy contamination.  Addition of another Gaussian line at 1402\angstrom{} improved the fit further to $\chi^2/dof=123.6/106$. We also accounted an emission feature near $1653\angstrom{} (7.5\ev{})$ by addition a Gaussian line, which further improved the fit to $\chi^2/dof = 111.44/102$. This emission feature is most likely due to \ion{He}{2} and \ion{O}{3}] \citep{2001AJ....122..549V}. We list the best-fit parameters  in Table~\ref{tabuvpar}, and show the unfolded spectra, best-fit model and the deviations of the data from the model in Figure~\ref{figfuvfit}. The photon indices $\Gamma_{\rm{FUV}}=2.4_{-0.6}^{+0.2}$ (obs1) and $2.8_{-2.2}^{+0.3}$ (obs2) of the FUV power law ($N_{E}\propto E^{-\Gamma_{\rm{FUV}}}$) correspond to the spectral indices $\alpha_{\lambda}=0.6^{+0.6}_{-0.2}$ (obs1) and $0.2^{+2.2}_{-0.3}$ (obs2)  since $\Gamma_{\rm{FUV}} = 3-\alpha_{\lambda}$ with $f_{\lambda} \propto \lambda^{-{\alpha}}$. { These spectral indices are consistent with the population mean and the standard deviation ($\alpha_{\lambda} = 1.1\pm0.8$) of the observed spectral indices for other AGN listed in Table~4 of \citealt{2002ApJS..143..257K}.}


\begin{table}
	\centering
	
	\caption{The best-fit FUV parameters of NGC~1566 with 90\% errors. The emission and absorption lines are identified from \cite{2001AJ....122..549V} and \cite{1999MNRAS.310..289O}, respectively.}
	\label{tabuvpar}
	\begin{tabular}{cccc}
		\hline\hline
		Element&	Parameter   &August 2018    &October 2018\\ 
		   &                &{ obs1}     &{ obs2}\\
		
		\hline
&  const.		    &$1.20\pm0.03 $  &$1.02\pm0.05$\\
		\\\
	
		&$E(B-V)$	&$0.0134$ (f)&	$0.0134$ (f) \\
		\hline
		\multicolumn{4}{c}{Absorption line}\\
	\ion{Ni}{2}	&$\lambda$ (\rm{\AA})	&$1744.4^{+5.9}_{-6.4}$	&  \\
	                      &FWHM$^a$		&$283.8^{+4407.7}_{-116.6}$	&  \\
	                     	&$\tau$		&$\geq0.24$ & \\
			\hline
				\multicolumn{4}{c}{Emission lines}\\
		\ion{He}{2}+\ion{O}{3}]  &$\lambda$ (\rm{\AA})	&$1651.5^{+6.7}_{-7.2}$	& $1651.5^{*}$ \\
	                &FWHM$^a$	&$7186^{+3800}_{-4152}$	& $7186^{*}$ \\
		            &flux$^b$					&$0.6\pm0.3$	&$0.10^{+0.10}_{-0.09}$ \\
		            \\\
		\ion{Si}{4}+\ion{O}{4}] &$\lambda$ (\rm{\AA})	&$1402.9^{+9.5}_{-7.5}$	& \\
	                &FWHM$^a$	&$\leq 6367.2$	& \\
		            &flux$^b$					&$0.32\pm0.17$	&\\
		\\\
		\ion{C}{4}&$\lambda$ (\rm{\AA})	&$1554.8^{+1.1}_{-1.2}$	&$1554.8^{*}$ \\
	               	&FWHM$^a$		&$6311.3^{+794.3}_{-787.4}$	&$6311.3^{*}$ \\
	            	&flux$^b$	    &$2.2\pm0.2$			&$1.3\pm0.2$ \\
		\hline
		\multicolumn{4}{c}{FUV continuum}\\
		&$\Gamma_{\rm{FUV}}$	&$2.4^{+0.2}_{-0.6}$	& $2.8^{+0.3}_{-2.2}$ \\
		&flux$^c$	&$1.76\pm0.07$	&$0.50\pm0.03$ \\
		\hline
		$\chi^2/dof$						&&	&$111.4/102$  \\
		\hline\hline									
	\end{tabular}
\\
    $^a$ FWHM in $\km{}\s^{-1}$\\
	$^b$ Flux in $10^{-12}$ \ergs{}\cm$^{-2}$\s$^{-1}$\\
	$^c$ Flux in 1300--1800\angstrom{} in 10$^{-11}$ \ergs{}\cm$^{-2}$\s$^{-1}$\\
\end{table}

\subsection{Joint FUV/SXT spectral analysis}

\begin{figure*}
	\centering
\includegraphics[scale=0.42]{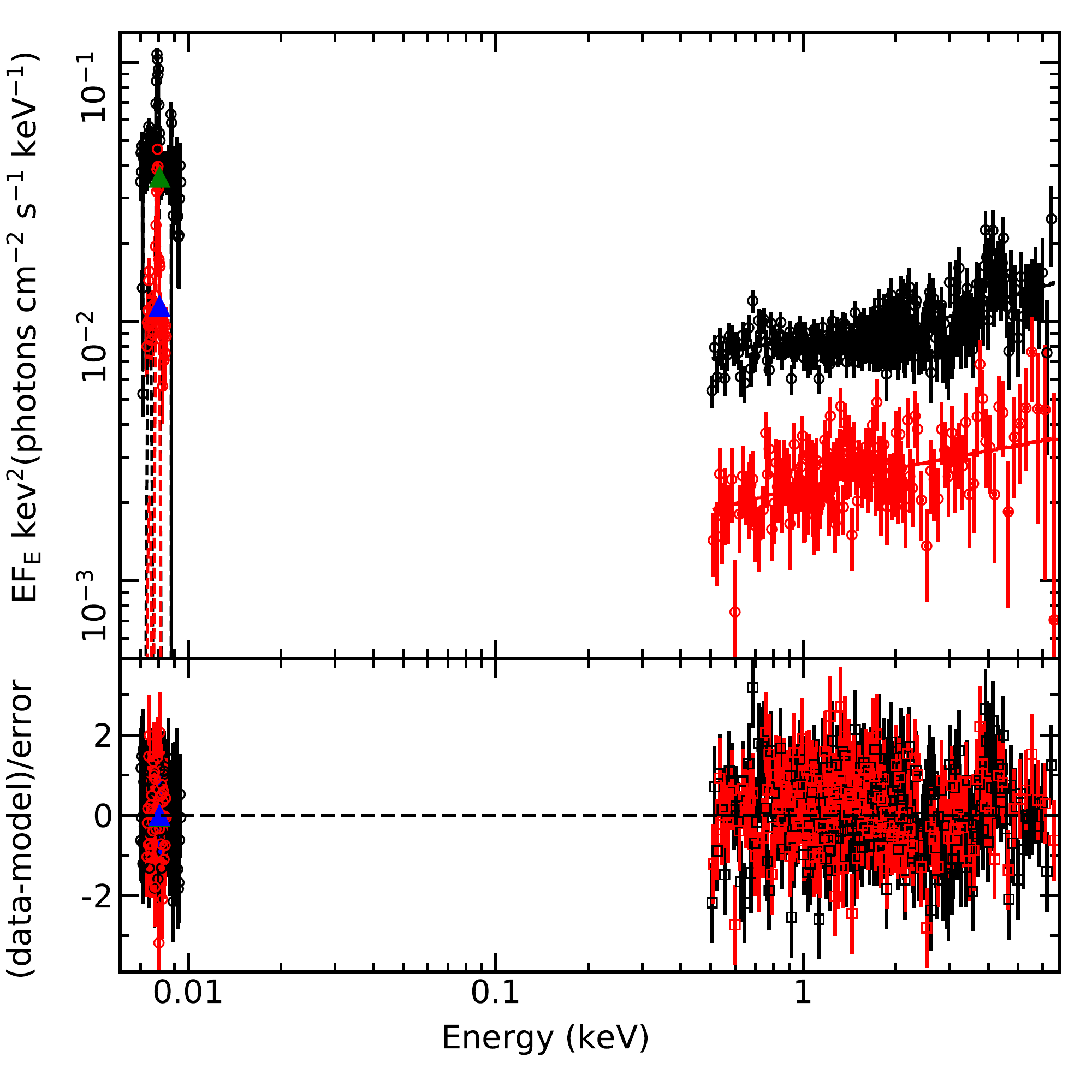}
\includegraphics[scale=0.42]{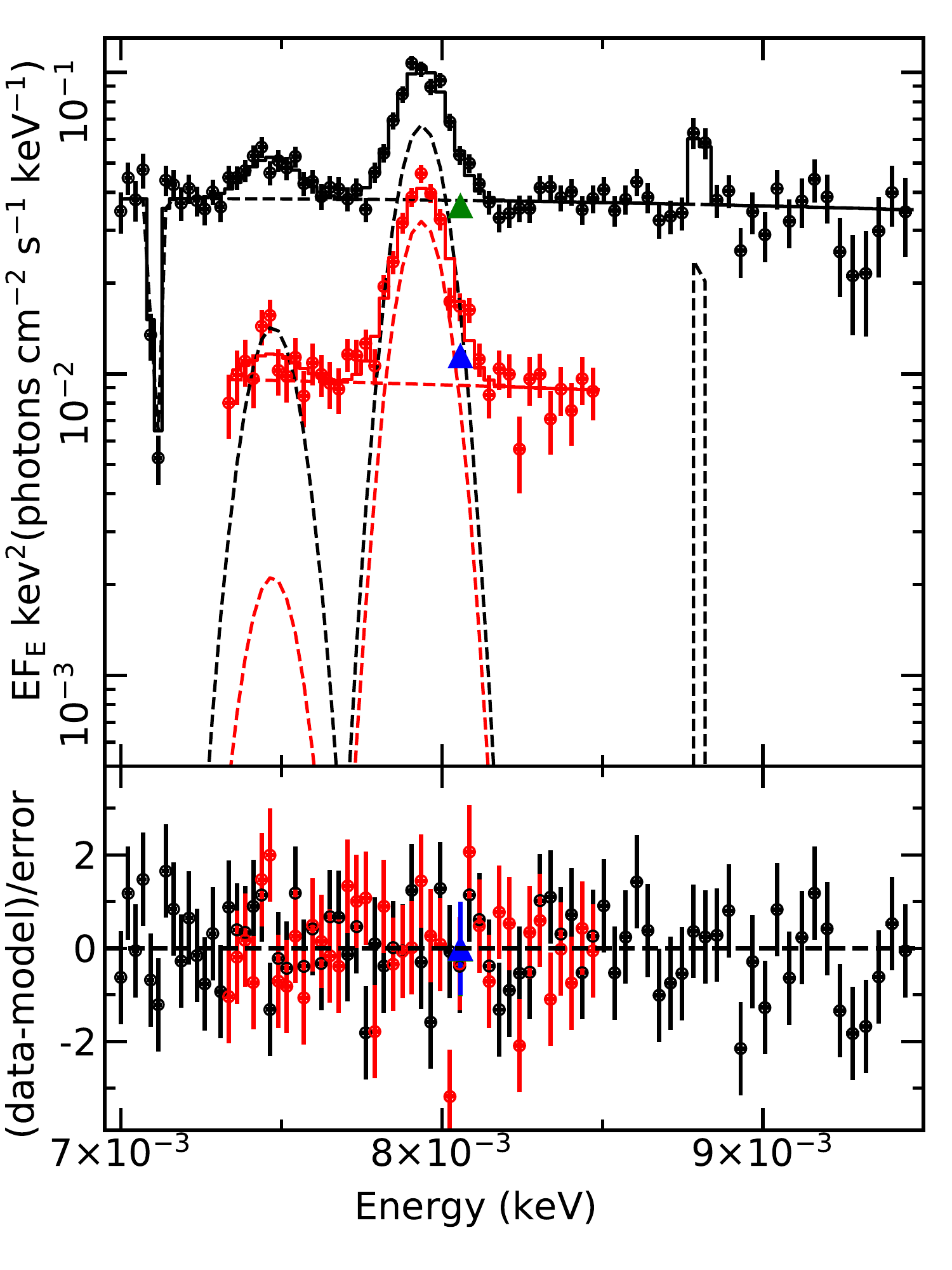}

\caption{{\it Left:} The broadband FUV/X-ray SEDs of NGC~1566 modeled with {\textsc{redden$\times$gabs$\times$tbabs$\times$(optxagnf+zgauss+zgauss+zgauss)}} model. The FUV-G1 and SXT data of obs1 and obs2 are shown in black and red colors, respectively. The F154W data of obs1 and obs2 are shown in green and blue colors, respectively. {\it Right:} A zoomed version of the left panel to show the FUV spectra, models, and residuals, clearly (here x-axis is plotted in the energy therefore position of the emission/absorption lines are changed with respect to Figure~\ref{figfuvfit}). }
\label{figfuvxrayfit}
\end{figure*}

We investigated the FUV/X-ray spectral variability of NGC~1566 by fitting the FUV and the X-ray data from the two \astrosat{} observations jointly. We fixed the parameters of FUV emission and absorption lines, and the relative normalization constants between the grating and broadband filter data at their best-fit values obtained earlier (see Table~\ref{tabuvpar}). We removed the {\textsc{zpowerlaw}} model which was used to fit the FUV continuum, and used the {\textsc{optxagnf}}~\citep{2012MNRAS.420.1848D} model to fit the FUV continuum, X-ray power-law and the soft X-ray excess component. The main parameters of the {\textsc{optxagnf}} model are, mass of the black-hole ($M_{BH}$), comoving distance to the source ($D$ in \mpc{}), dimensionless accretion rate ($\log{\dot{m}=L/L_{Edd}}$), dimensionless black-hole spin ($a$), inner and outer disk radii ($r_{cor}$ and $r_{out}$), temperature ($kT_{w}$) and optical depth ($\tau_{w}$) of the warm corona, X-ray power-law photon-index ($\Gamma$) and fraction of the power below $r_{cor}$ which is emitted in the hard comptonization component ($f_{pl}$). 
{ We fixed the black-hole mass at $M_{BH}=8.32\times10^{6} M_{\odot}$~\citep{2002ApJ...579..530W} following the previous studies \citep[see][]{2019MNRAS.483L..88P, 2020MNRAS.498..718O,2021MNRAS.507..687J}, the black-hole spin $a=0$ for a slowly spinning black-hole \citep{2019MNRAS.483L..88P,2021MNRAS.507..687J},} and the outer disk radius $r_{out}=10^{5}R_g$, where $R_g = GM_{BH}/c^2$ is the gravitational radius. The comoving distance of the source is  uncertain, the reported distance ranges from 5.5\mpc{}~\citep{2014MNRAS.444..527S}
to 21.3\mpc{}~\citep{2019MNRAS.487.2797E,2004MNRAS.350.1511O}. 
{ In our analysis, we adopt the comoving distance $D = 21.3\mpc{}$ based on the Hubble constant $H_{0} = 70.3\km{}\s^{-1}\mpc^{-1}$, which is also consistent with the most recently reported Tully-Fisher distance of the source $16.9^{+7.5}_{-4.1}\mpc{}$ \citep{2019MNRAS.487.2797E} }. 
We used {\textsc{tbabs}} model for the Galactic absorption in the X-ray band only and {\textsc{redden}} model for the Galactic reddening in the FUV band only. We tied the parameters of \textsc{optxagnf} for the SXT spectra with those of the FUV spectra for each observation. We varied the mass accretion rate, inner disk radius, photon-index and fraction of the X-ray power-law component independently for each observation. We also fixed the warm corona temperature at the best-fit value derived from the SXT data using the {\textsc{nthcomp}} model earlier. We tied the optical depth across the two observations and allowed it to vary. As before, we applied 3\% systematic error and the gain correction. While fitting we found that the $f_{pl}$ parameter is 1 for the obs2 as we did not detect the soft X-ray excess component in this observation. We therefore fixed this parameter at 1 for the obs2. This did not change other parameters, and resulted in an acceptable fit with  $\chi^2/dof=522.7/518$. We list the best-fit parameters in Table~\ref{tabuvxraypar}. We show the UV/X-ray spectral data, the best-fit model and the deviations of the data from the model in Figure~\ref{figfuvxrayfit}.

\begin{table*}
	\centering
	\caption{The best-fit parameters of NGC~1566 derived from the broadband UV/X-ray SED.}
	\label{tabuvxraypar}
	\begin{tabular}{cccccc}
		\hline\hline
		Model&	Parameter       &\multicolumn{2}{c}{\astrosat{}} &\multicolumn{2}{c}{\xmm{}}\\
		    &                      &August 2018 &October 2018   &June 2018  &November 2015\\
		\hline
		{\textsc{optxagnf}}	&$\log{\dot{m}}$ 	&$-1.81^{+0.02}_{-0.07}$	&$-2.44^{+0.03}_{-0.04}$ &$-1.29^{+0.01}_{-0.01}$ & $-2.91^{+0.01}_{-0.01}$ \\
		                    &$a$ 	&0 (f)	&0 (f) &0 (f) & 0 (f)\\
		                     &$r_{cor} (R_g)$		&$52^{+7}_{-7}$	&$38.9^{+5.8}_{-1.9}$ &$45^{+8}_{-7}$ &$25.9^{+0.9}_{-0.8}$\\
		                    &$kT_{w} (\kev{})$    &0.15 (f)    & -&$0.51^{+0.05}_{-0.05}$&\\
		                    &$\tau_{w}$              &$\geq 25 (3\sigma)$  &- &$9.5^{+0.5}_{-0.5}$ &\\
		                    &$\Gamma$                 &$1.63^{+0.07}_{-0.02}$ &$1.78^{+0.07}_{-0.03}$ &$1.70^{+0.02}_{-0.02}$ &$1.86^{+0.01}_{-0.01}$\\
		                    &$f_{pl}$                   & $0.979^{+0.003}_{-0.015}$  &1 (f)  & $0.49^{+0.03}_{-0.03}$    &1 (f)\\
		                    \\\
		 {\textsc{relxill}} &$N_{rel} (10^{-5})$  &&&$4\pm1$   &\\
		 {\textsc{xillver}} &$N_{xill} (10^{-5})$ &&&$5.6\pm1.3$    &$1.2\pm0.2$\\

		\hline
		{\textsc{flux}} &$f_{disk}^a$   &10.63   & 3.01  &40.46   &1.35\\
		                &$f_{SXE}^b$    &0.28    & -  &2.07    &-\\
		                &$f_{PL}^c$     &3.23    & 0.85  &5.83 &0.23\\
		                \hline
	$\chi^2/dof$						&&&$522.7/518$ &$246.8/238$ &$243.9/221$ \\
		\hline\hline									
	\end{tabular}
	\\$^a$ = The disk flux in $1~\mu-50\ev{}$ band in $10^{-11}\ergs{}\cm^{-2}\s^{-1}$\\
	$^b$ = The soft X-ray excess flux in $0.5-2\kev{}$ band in $10^{-11}\ergs{}\cm^{-2}\s^{-1}$\\
	$^c$ = The power-law flux in $2-10\ev{}$ band in $10^{-11}\ergs{}\cm^{-2}\s^{-1}$\\
\end{table*}

\subsection{\xmm{} UV/X-ray spectroscopy}
\begin{figure}
    \centering
    \includegraphics[scale=0.54]{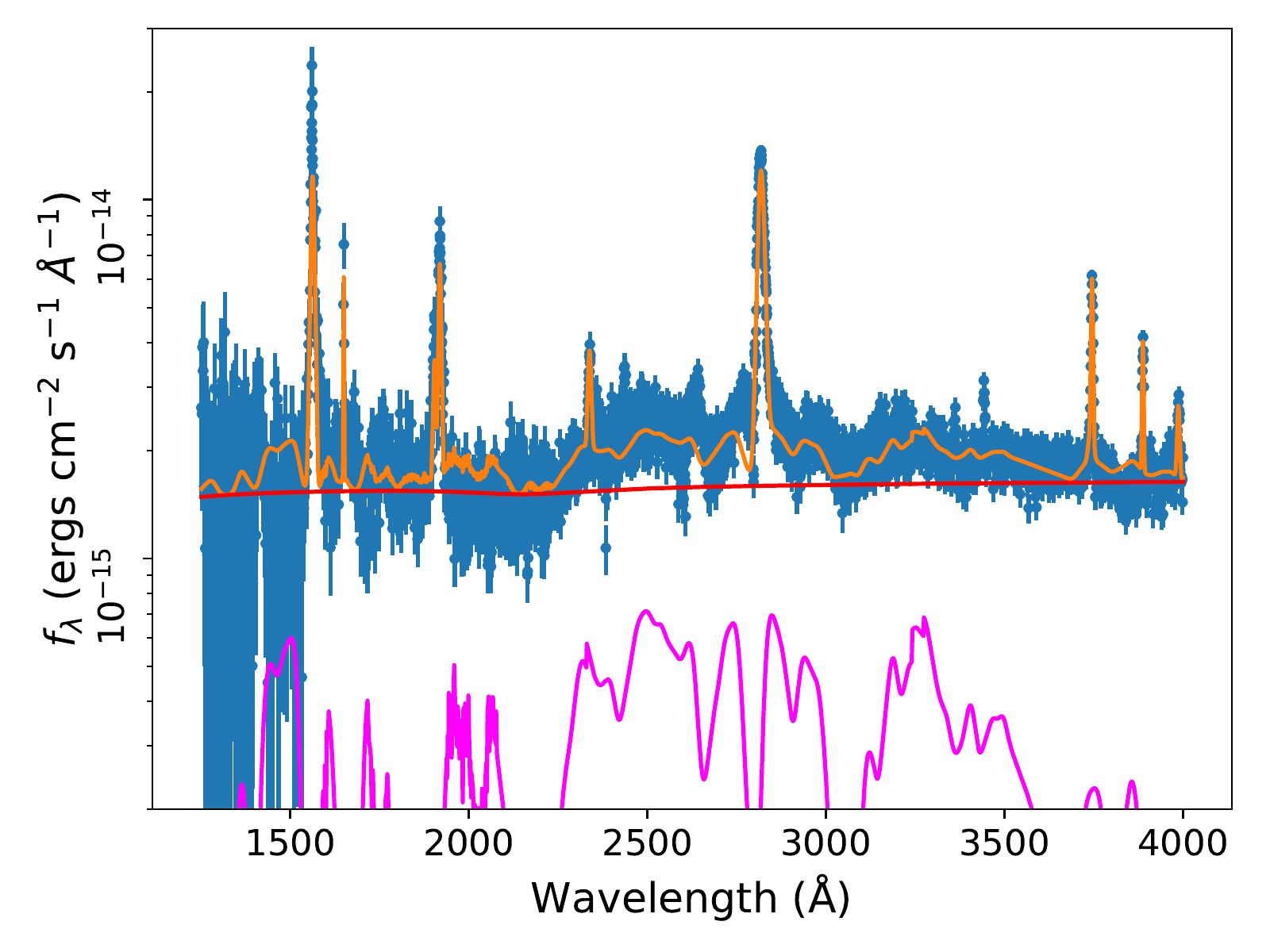}
    \caption{The \hst{} spectrum of NGC~1566 modeled with a {\textsc{powerlaw}} modified by the Galactic reddening (red), Gaussian line models for emission lines, and Fe~II templates (magenta)}
    \label{fighst_spec}
\end{figure}

In order to investigate the broadband UV/X-ray spectral evolution of NGC~1566, we also analyzed two \xmm{} observations of June 26, 2018 (Obs.ID: 0800840201) when the source was at the peak of the outburst, and November 05, 2015 (Obs.ID: 0763500201) when the source was at a low flux state before the outburst. We processed the EPIC-pn data in the same way as described in \cite{2019MNRAS.483L..88P}, and derived the source and background spectra. Additionally, we processed the OM data using the {\textsc{omichain}} task and generated images ready for the aperture photometry. The optical/UV data are contaminated by the host-galaxy emission, we therefore used the OM data below $3000$\angstrom{} only to avoid stronger host galaxy contamination. We performed the aperture photometry using {\textsc{omsource}} task for UVW2 ($\lambda_{\rm{eff}} = 2120$\angstrom{}, $\Delta{\lambda} = 500$\angstrom{}) and UVM2 ($\lambda_{\rm{eff}} = 2310$\angstrom{}, $\Delta{\lambda} = 480$\angstrom{}) filters of 2 observations. Following \citealt{2018A&A...619A..20M}, we extracted the source counts from a circular region of 13 pixels (1 pixels $= 0.953\arcsec$) radius centered on the source and background counts from a similar size of annular region outside the source but inside the host-galaxy. We calculated the net background subtracted source count rates for the 2 filters. We corrected the net source count rates for the Galactic extinction using the CCM extinction law \citep{1989ApJ...345..245C} with a color excess of $E(B-V)=0.0134$ and the ratio of total to selective extinction of $R_V = A_V/E(B-V) = 3.1$, where, $A_V$ is the extinction in the $V$ band. { Further, the optical/UV data are also contaminated by the BLR/NLR emissions, which cannot be corrected directly from the broadband filter data. We therefore used an \hst{} spectrum of NGC~1566 in the $1250-4000$\angstrom{} band to estimate the fractional contribution of the BLR/NLR emission to the total AGN flux. We used the G270H grating data of the Faint Object Spectrograph (FOS)/\hst{} observed on 08 February 1991 (ObsID: y0h70207t) with a total exposure time of 2\ks{}. We obtained the G270H grating/\hst{} spectrum from the MAST portal\footnote{\url{https://mast.stsci.edu/portal/Mashup/Clients/Mast/Portal.html}}.} 
We modeled the \hst{} spectrum in {\textsc{sherpa}}~\citep{2001SPIE.4477...76F}. We used a {\textsc{powerlaw}} modified by the Galactic reddening ({\textsc{ccm}}) model for the UV continuum, Gaussian line models for emission lines, and a Fe~II template convolved with the Gaussian-smoothing model {\textsc{gsmooth}}. {We fixed the $\sigma_{\rm{Fe~II}}$  of the smoothing component at the $\sigma$ width of the broad emission lines corresponding to FWHM $\sim 2300\km{}\s^{-1}$}. Following \citealt{2015A&A...575A..22M}, we also added the Balmer continuum function 
and a host-galaxy bulge template~\citep{1996ApJ...467...38K}, which did not change the fit. We therefore did not use these components further. The \hst{} spectrum and models are shown in Figure~\ref{fighst_spec}. 
From the best-fit model of the \hst{} spectrum, we removed the reddening 
component, and derived the de-reddened total model and {\textsc{powerlaw}} model. We treated the de-reddened {\textsc{powerlaw}} model as the UV continuum. We used the effective area of each filter, and calculated the count rates of the de-reddened {\textsc{powerlaw}} ($f_{cont}$) and total model ($f_{total}$) in the filter's band. Using these contributions, we finally calculated the fractional contributions of the non-continuum components in the total spectrum (i.e.,$1-(f_{cont}/f_{total})$). We found these fractions to be $\sim 23\%$ for UVW2 and $\sim 26\%$ for UVM2. We subtracted these fractions from the extinction corrected count rates and derived the intrinsic count rates of the source. We wrote these count rates in a 
{OGIP compliant} spectral file generated using {\textsc{om2pha}} task.

\begin{figure}
   \centering
   \includegraphics[scale=0.42]{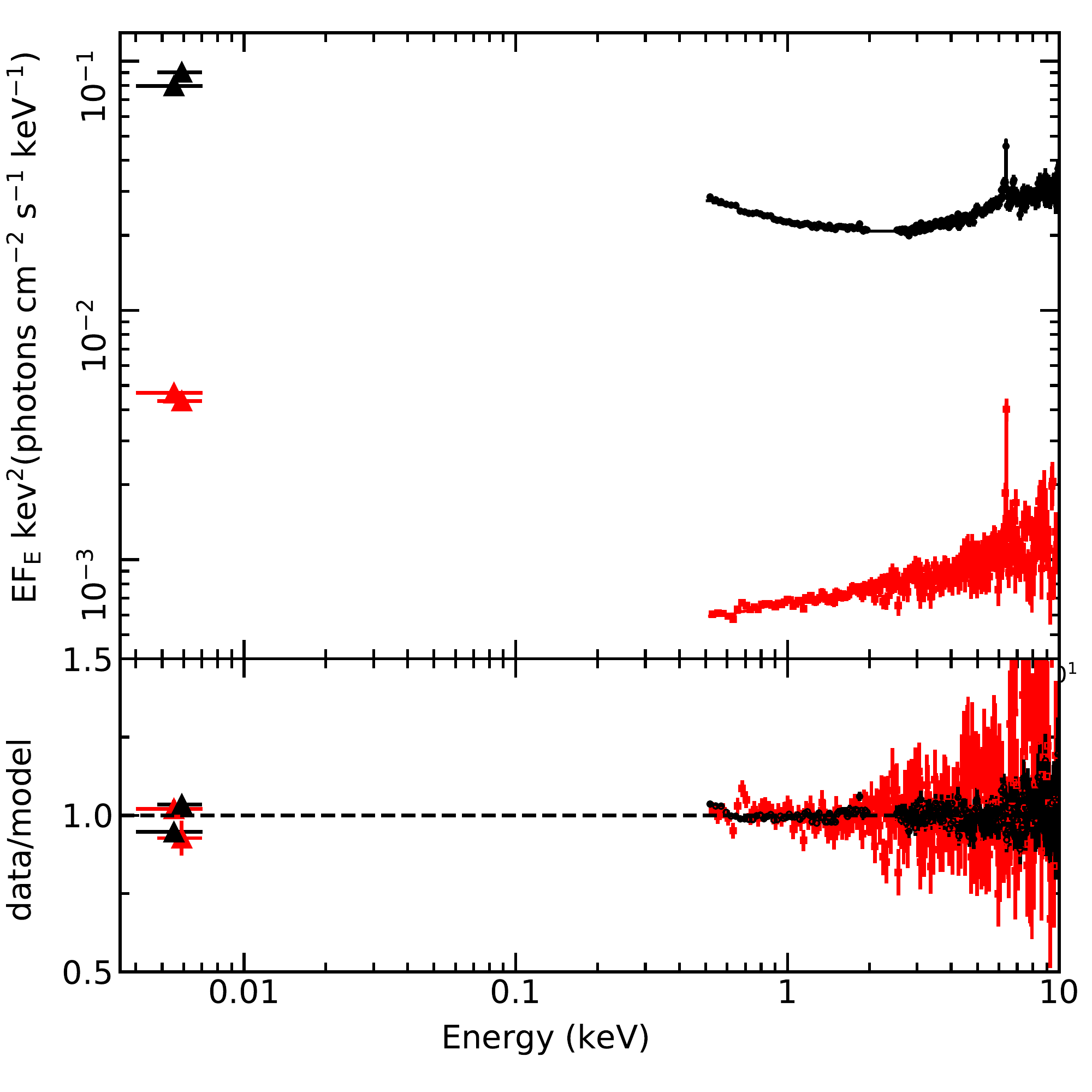}
    \caption{UV to X-ray SEDs of NGC~1566 derived using 2015 (red) and 2018 (black) \xmm{} observations. The UV data are shown in triangles for both observations and the EPIC-pn data for 2015 and 2018 are shown in open squares and open circles, respectively. The bottom panel shows the residuals in terms of data-to-model ratio.}
    \label{figxmm_seds}
\end{figure}


As before, we used the {\textsc{optxagnf}} model to fit the X-ray power-law, the soft X-ray excess, and the UV continuum, simultaneously. We modeled the distant and blurred reflection features using {\textsc{xillver}} and {\textsc{relxill}} components, respectively. We fixed the parameters of the reflection models at their best-fit values derived by \cite{2019MNRAS.483L..88P} such that the emissivity index ($q=3$), iron abundance ($A_{Fe}=3$), inclination angle ($\theta=10\degree$), ionization parameter ($\log{\xi = 2.4}$), spin of the SMBH ($a=0$), and high energy cut-off ($E_{cut}=167\kev{}$). We also fixed the reflection fraction $R=-1$ for both the reflection models. We also added the additional Gaussian line at 6.85\kev{} reported by \cite{2019MNRAS.483L..88P}, and fixed its parameters. We added $1\%$ systematic in the joint fitting. The best-fit resulted in $\chi^2/dof = 246.8/238$ for 2018 data. Similarly, we also fitted the 2015 data using the {\textsc{optxagnf}} and {\textsc{xillver}} models. We did not require the blurred reflection component for this dataset. While fitting we found that the $f_{pl}$ parameter of the {\textsc{optxagnf}} model is close to 1, which is expected due to the negligible soft X-ray excess at this epoch. We therefore fixed this parameter at 1 which did not change the fit. The fit resulted in $\chi^2/dof=243.9/221$ for 2015 data. We list the best-fit parameters and the fluxes of the accretion disk, soft X-ray excess, and the power-law components for both the \xmm{} observations in Table~\ref{tabuvxraypar}. We separately calculated the soft X-ray excess flux for 2015 data using the {\textsc{zbbody}} model, which resulted $kT_{bb} = 0.16\pm0.02\kev{}$ and the $0.5-2\kev{}$ flux $8.5\pm3.5 \times10^{-14}\ergs{}\cm^{-2}\s^{-1}$.

\subsection{Blurred reflection and the soft X-ray excess emission}

{ We investigated if the strong soft X-ray excess emission present in the 2018 \xmm{} can be modeled with the blurred reflection from the accretion disk as a result of the coronal X-ray  illumination onto the disk. We  used a high density blurred reflection model, {\textsc{relxilld}} \citep{2016MNRAS.462..751G}, to fit the soft X-ray excess emission, broad iron line, and the power-law continuum. We allowed all the parameters to vary freely. We found statistically good fit only for a highly spinning black-hole ($a>0.6$) and strong reflection ($R\sim 1.3$), which is inconsistent with the black-hole spin of $a\leq0.2$ and reflection fraction of $R<0.2$ derived from the broad iron line alone \citep[see][]{2019MNRAS.483L..88P, 2021MNRAS.507..687J}. Generally  the smooth soft X-ray excess component when described as the blurred reflection model requires  maximum black-hole spin  \citep[see][]{2018MNRAS.479..615M,2019MNRAS.489.3436J}. 
To further investigate, we also fitted 3--10\kev{} EPIC-pn data from the June 2018 observation  with a {\textsc{zpowerlaw}} for the X-ray continuum, {\textsc{xillver}} for the distant reflection, and {\textsc{laor}} line for the broad iron line. We fixed the line energy at 6.4\kev{}, inclination angle at 10\degree as before, and normalization of the {\textsc{xillver}} at the best-fit value derived above. The fit resulted in $\chi^2/dof = 155/165$. The best-fit parameter of the {\textsc{laor}} line  are the inner disk radius $R_{in} = 15^{+18}_{-9} R_g$, emissivity index $q > 2.4$, and the equivalent width $EW = 45^{+20}_{-17}\ev{}$. Thus, the inner disk radius is also consistent with a low spinning black-hole, and a small equivalent width of the line suggests that the blurred reflection is not strong in this source. Therefore, the strong soft X-ray excess emission observed in the June 2018  is unlikely to arise due to the blurred reflection.}

\begin{figure*}
   \centering
\includegraphics[scale=0.82]{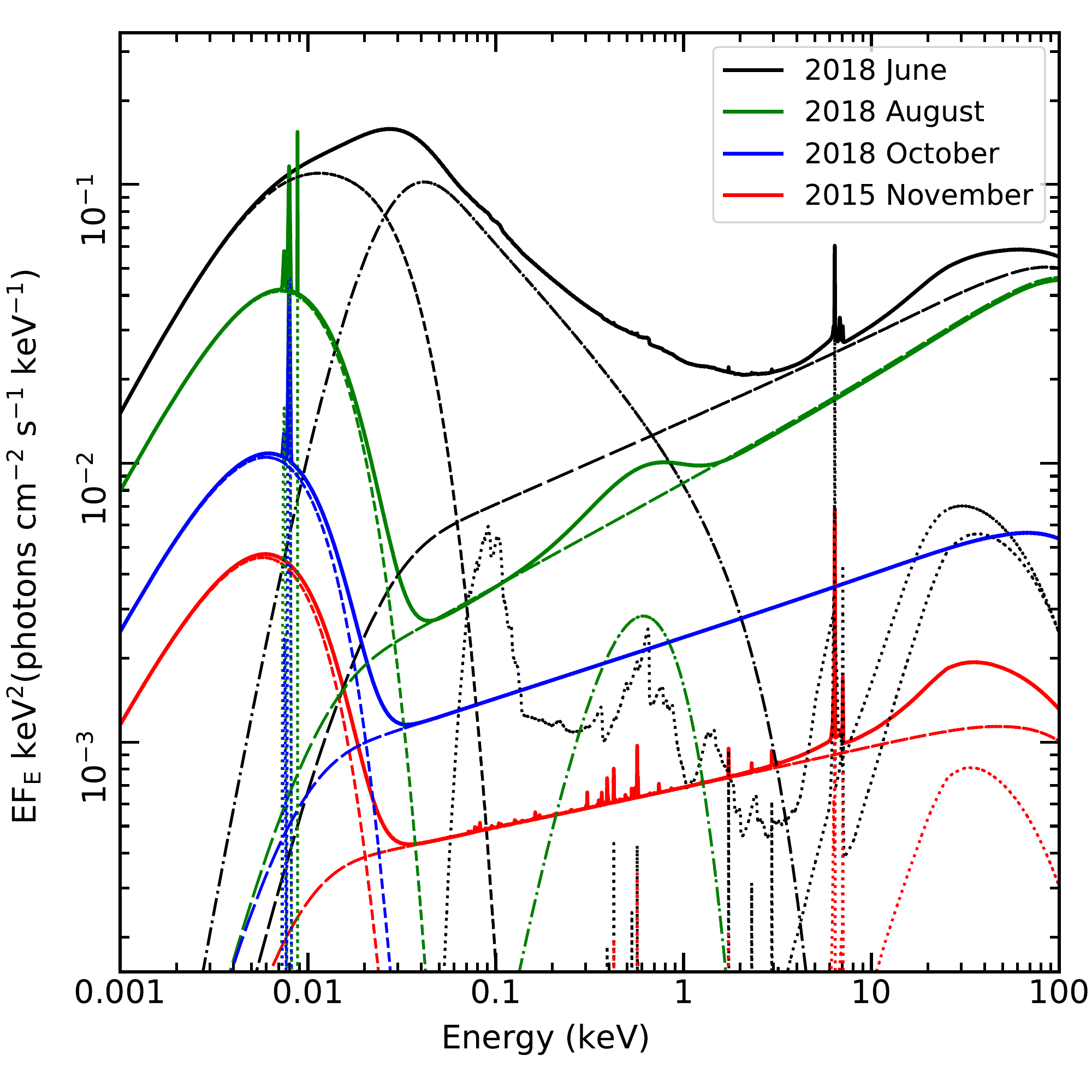}
    \caption{The best-fit UV/X-ray models for 2018 June (black), 2018 August (green), 2018 October (blue), and 2015 November (red) derived using 2 \xmm{} and 2 \astrosat{} data. The accretion disk (short dashed lines), X-ray power-law (long dashed lines), and soft X-ray excess (dash-dotted lines) components for each epoch are shown with the same colors. The reflection models and the UV emission lines are shown in the dotted lines.} 
    \label{figxmm_astrosat_SEDmodels}
\end{figure*}

\section{Results and Discussion}
\label{sec_results}

We analyzed two observations of NGC~1566 in the FUV and X-ray bands performed by \astrosat{} in August and October 2018 during the  declining phase of the 2018 outburst. We also analyzed two \xmm{} observations performed before the outburst in 2015 and during the peak of the outburst in June 2018. Using these four UV/X-ray observations, we  investigated the changes in the primary spectral components -- the accretion disk, the soft X-ray excess, and the X-ray power-law during the 2018 outburst. The main results of our analysis are as follows.
\begin{enumerate}[label=(\roman*)]
\item The FUV spectra of NGC~1566 derived using the FUV-G1 and F154W data consists of the FUV continuum, the emission lines of \ion{C}{4}, \ion{Si}{4}+ \ion{O}{4}] and \ion{He}{2}+\ion{O}{3}]. 
{ The width of these lines (FWHM$\sim 4000-7000{\rm~km~s^{-1}}$)} are consistent with their origin in the broad-line region.
    \item The observed flux in the FUV (1300--1800\angstrom{}) and X-ray ($0.5-6.8\kev$) bands decrease by a factor of $\sim 3$ in $\sim 70$ days during the declining phase from August to October 2018 (see Table~\ref{tabobs}).    
    \item The reduction in the ionising flux also led to reduced strengths of broad UV emission lines, the  \ion{C}{4}$\lambda1549$  line varied by a factor of $\sim 2$.  
    \item The UV/X-ray broadband SEDs of NGC~1566 are well described by  accretion disk component superimposed with emission lines, warm Comptonisation responsible for the soft X-ray excess, the X-ray power law, distant and blurred reflection \citep[see also][]{2019MNRAS.483L..88P, 2019MNRAS.483..558O, 2013ApJ...770..157K}. 
    \item All the UV/X-ray spectral components strongly varied during the outburst. The accretion disk continuum emission increased by a factor $\sim 30$ from pre-outburst in 2015 to peak outburst in June 2018, then declined by a factor of $\sim 13$ in four months.
    \item The soft X-ray excess component was extremely weak, almost undetectable in 2015. This component increased by a factor of $\sim 243$ during the peak outburst, then decreased by a factor of $\sim 7$ in August 2018 and again became undetectable in October 2018. 
    \item The X-ray power-law component increased by a factor of $25$ from 2015 to June 2018, decreased by factors of $\sim 1.8$ in August 2018 and $\sim 6.8$ in October 2018. The X-ray power-law was steeper in the lowest flux states and appears to have flattened at high flux states.
    \item The bolometric luminosity relative to the Eddington luminosity ($L/L_{Edd}$) increased from $\sim0.1\%$ (2015) to $\sim 5\%$ (June 2018), then decreased to $\sim 1.5\%$ (August 2018) and $\sim 0.3\%$ (October 2018).

\end{enumerate}

 We summarize the spectral evolution of NGC~1566 in 
 Figure~\ref{figxmm_astrosat_SEDmodels} that shows the best-fitting UV to X-ray broadband models derived from the observations in June 2018 (black), August 2018 (green), October 2018 (blue), and November 2015 (red). The spectral components -- the accretion disk (dashed lines), soft X-ray excess (dot-dashed lines), X-ray power-law (long-dashed lines), and reflection components (dotted lines) are also shown in Figure~\ref{figxmm_astrosat_SEDmodels}. Clearly, NGC~1566 has shown one of the most extreme variability in its primary spectral components. In particular, the soft excess component dropped by a factor $> 45$ during the declining phase that is more than any other component. The broadband spectra of NGC~1566 is similar to that observed from other CL-AGN such as Mrk~1018 \citep{2018MNRAS.480.3898N} that also showed most extreme variability in its soft X-ray excess emission in $\sim 8$ years. Another CL-AGN Mrk~590 is also known to show drop in its soft excess emission by a factor of $\sim 20-30$ in 7 years while the  $2-10\kev$ continuum showed only $\sim10\%$ change in flux \citep{2012ApJ...759...63R}. Similarity of the observed spectral variability associated with the changing-look phenomena in these AGN points towards a fundamental process that is not well understood. 
 

\subsection{Variation in the accretion disk emission}
The far UV continuum in the 1300--1800\angstrom{} band observed with the \astrosat{}/UVIT and the near UV band observed with the \xmm{}/OM are consistent with the standard accretion disk around a black hole with $M_{BH}=8.32\times10^6M_{\odot}$. The accretion disk is found to contribute $0.07\%$ (2015), $2.15\%$(June 2018), $0.57\%$ (Aug 2018) and $0.16\%$ (Oct 2018) in the $1\micron$ to $50\ev$ band relative to the Eddington luminosity. The intrinsic disk Comptonization model {\textsc{optxagnf}} which we used to fit the broadband spectra suggests that the change in the disk flux is governed not only by the change in the accretion rate but also due to the change in the { size of the warm Comptonizing medium ($r_{cor}$)} 
from $\sim26r_g$ in 2015 to $50r_g$ in June-August 2018.  The disk flux increased by a factor of $\sim 30$ despite possible decrease in the emitting area of the standard disk from 2015 to 2018. The inner part of the disk appears to have been converted in the warm Comptonising medium giving rise to the soft X-ray excess in the high flux states. 
As pointed out by a number of authors (e.g. \citealt{2020ApJ...898L...1R, 2019ApJ...887...15W, 2018MNRAS.480.3898N}, the viscous time scale for accretion disks in AGN is much longer than the changing-look time scale in AGN. 
To compare the disk variability timescale, we calculate the accretion disk timescales at the inner radius of the disk obtained from the best-fit broadband spectral model.
The dynamical ($t_{dyn}$), thermal ($t_{th}$), and viscous ($t_{vis}$) timescales of the accretion disk are given by, \citep{2006ASPC..360..265C}
\begin{equation}
t_{dyn} = \bigg(\frac{r^3}{GM_{BH}}\bigg)^{1/2} 
\end{equation}
\begin{equation}
t_{th} = \frac{1}{\alpha}t_{dyn},
\end{equation}
and 
\begin{equation}
t_{vis} \approx \frac{1}{\alpha}\bigg(\frac{r}{h}\bigg)^2t_{dyn},
\end{equation}
\noindent
where, $r$ is radial distance in the disk,
$\alpha$ is the viscosity parameter, and $h$ is height of the disk. We used the results from the August 2018 observation and  calculated the accretion disk temperature of $\sim 3.6\ev{}$ for an inner disk radius of $\sim 50 R_g$, accretion rate of $\dot{m}\sim 0.015$, and a black-hole mass of $8.32\times10^{6}M_{\odot}$. We used this temperature to estimate the height-to-radius ratio of the disk $h/r = c_s/v_{\phi} \sim 4.4\times10^{-4}$ (where $c_s = \sqrt{kT/m_p}$ is the sound speed and $v_{\phi} = \sqrt{GM_{BH}/r}$ is the Keplarian velocity). Assuming $\alpha=0.1$, we finally calculated the dynamical, thermal, and viscous timescales $t_{dyn}\sim0.17$ days, $t_{th}\sim1.7$ days, and $t_{vis}\sim 2.4\times10^{4}$ years, respectively. The timescale for the observed outburst is much smaller than the viscous timescale but longer than the dynamical and thermal timescales at an inner radius of $50r_g$. On the other hand the estimated sound crossing time of $t_{s}\sim 50R_g/c_s \sim 1{\rm~year}$ is comparable to the changing-look time of NGC~1566. This suggests that the outburst is likely related with  pressure instabilities in the disk.

\subsection{Variations in the soft X-ray excess emission and its origin}
The soft X-ray component exhibited the most extreme variations during the outburst. This component is well described by thermal Comptonization of the accretion disk photons in a warm and optically thick corona. The strongest soft excess observed during the peak of the outburst is described by warm plasma with $kT_w=0.51\pm0.05\kev$, $\tau=9.5\pm0.5$. After about two months, the soft excess weakened with increased optical depth and decreased temperature. 
The variability of the soft X-ray excess component is larger than the disk and the power-law components. Therefore, it can neither be produced by the warm Componization of the disk photons without an intrinsic change in the warm Comptonising medium, nor by the illumination of the X-ray power-law component alone. The soft excess component was almost non-existent before the outburst and towards the end of the outburst, and it was at its maximum during the outburst peak.  The warm Comptonising region emitting the soft X-ray excess is characterized by the transition radius $r_{cor}$ between the disk and the warm corona, and below this radius the warm corona is thought to exist. It appears that this inner region is formed during the outburst on a time scale comparable to the sound crossing time at the inner radius of the standard disk. It also appears to have disappeared towards the end of the outburst on the sound crossing time. The increasing optical depth and decreasing temperature of the warm Comptonising medium during the declining phase, and finally the disappearance of the soft excess towards the end of the outburst imply that the warm Comptonising medium most likely converted back to the standard inner disk material.

\subsection{Variation in the X-ray power law \& Thermal Comptonization}

We found that the X-ray power-law flux increases with the increasing disk and soft X-ray excess fluxes (see, Table~\ref{tabuvxraypar}). Similar correlations have been found in other AGN also -- IC~4329A~\citep{2021arXiv210413031T}, PKS~0558-504~\citep{2013MNRAS.433.1709G}, NGC~7469~\citep{2001AdSpR..28..295N}, and interpreted in terms of thermal Comptonization of the seed photons in the hot corona. The number of scattering in the hot corona increases with the number of increasing seed photons which increases the Comptonized X-ray flux. It is possible that the accretion disk and the soft X-ray excess both components provide the seed photons for thermal Comptonizatoion in the hot corona. The increasing seed photons cools the corona which results in a steeper photon-index (see, \citealt{nthcomp1},\citealt{nthcomp2}). However, the X-ray power-law photon-index is not correlated with the disk and/or soft X-ray excess flux in our case. This implies that the outburst also caused intrinsic change in the hot corona affecting its optical depth and temperature. \nustar{} observations performed during the outburst should reveal possible intrinsic change in the hot corona.

\subsection{Spectral Transition}

The physical mechanism used to explain such drastic variability include variable obscuration, TDE events, and disk instabilities. The variable obscuration due to changing clouds along the line-of-sight block the direct AGN emission which causes a formal classification change of the source \citep{2002MNRAS.329L..13G, 2003MNRAS.342..422M}. In the case of NGC~1566, the variable obscuration mechanism is ruled out as the low flux X-ray spectra are free from intrinsic absorption. 
\cite{2019MNRAS.483..558O} proposed that increasing luminosity may be as a result of sublimation of the dust in the line-of-sight which previously obscured the part of the broad line region. The greater luminosity of the direct emission from the central regions can also increase the intensity of the Balmer lines. In our analysis, we found that fluxes of the broad emission lines are correlated with the X-ray power-law  and the FUV continuum flux, but we did not find additional reddening by the dust at the source-redshift. Thus, the obscuration of the BLR due to the dust along the line-of-sight is unlikely to be responsible for the multi-wavelength flux change.

TDEs are also thought to produce such events. For example, \cite{2020ApJ...898L...1R} used TDE mechanism to explain the changing look behavior of 1ES1927+65. In case of NGC~1566, outbursts are observed repeatedly \citep{1986ApJ...308...23A}. Even, during the declining period of the 2018 outburst, \cite{2020MNRAS.498..718O} observed 3 additional bursts in $\sim 1$ year. The theoretical rate of TDEs is { too low} $\sim 10^{-4}-10^{-5}$ galaxy$^{-1}$ year$^{-1}$ (See, \citealt{2016PASJ...68...58K}, \citealt{2015JHEAp...7..148K}, and references therein) to produce such repetitive events. Also, the X-ray spectra of NGC~1566 are harder than a typical X-ray spectrum of TDEs \citep{2015JHEAp...7..148K, 2017AN....338..256K}. Thus, the TDEs are unlikely to be responsible for the burst activities in NGC~1566.

NGC~1566 showed extreme UV/X-ray flux and spectral variability during the 2018 outburst (see Figure~\ref{figxmm_astrosat_SEDmodels}). The emergence of the soft X-ray excess component during the outburst requires major structural changes in the innermost regions. Thus, the observed spectral changes are not merely due to minor changes in the physical conditions of the existing emitting regions but the transition from a negligible soft excess state to a strong soft excess state and back to the negligible soft excess state requires formation and disappearance of the warm Comptonising medium. The accretion rate is found to increase with the onset of the outburst, which results in the increased energy generation in the innermost disk. This increased energy generation most likely could not all be released in the form of radiation, and the trapping of photons raises the  temperature and pressure, and most likely decreases the density or the optical depth of the inner disk,  thus transforming the innermost accretion disk to the warm Comptonization medium.  This in turn gave rise to the observed soft X-ray excess emission via thermal Comptonization in the warm, optically thick innermost disk. The final configuration is likely the same as already hypothesized in the intrinsic disk Comptonized models ({\textsc{optxagnf}}; \citealt{2012MNRAS.420.1848D} and {\textsc{agnsed}}; \citealt{2018MNRAS.480.1247K}). Unlike the narrow-line { Seyfert~1} galaxies, the soft X-ray excess from NGC~1566 does not seem to persist for long time but lasted only during the outburst. During the declining phase, the soft excess emitting region appears to have disappeared, most likely  transforming  to innermost disk. This transition appears to occur at a few percent of the Eddington rate below which there is no soft excess emitting region but only a standard disk and hot corona exist. It is thus likely that NGC~1566 makes transitions between a cold disk and  a cold+warm disk. 


 The changing look phenomena appears to occur in a number of AGN including NGC~1566 at a few percent of the Eddington rate e.g., Mrk~1018 \citep{2018MNRAS.480.3898N}, Mrk~590 \citep{2014ApJ...796..134D}, NGC 2617 \citep{2014ApJ...788...48S}, and {SDSS J0159+0033} \citep{2015ApJ...800..144L}. At  Eddington fraction of a few percent, black hole X-ray binaries (BHB) are known to make high/soft to low/hard spectral transition. Hence, similar spectral transition has been suggested for the changing-look events in Mrk~1018 \citep{2018MNRAS.480.3898N} and in a sample of changing-look quasars \citep{2019ApJ...883...76R}. However, there are subtle differences. In the high/soft to low/hard spectral transition of BHB, the disk emission diminishes and the X-ray power law component strengthens and flattens, while the disk emission, soft excess and power-law component all vary in a correlated manner in CL-AGN. The X-ray power law does not appear to flatten in the low flux states of CL-AGN. There does not seem to be a spectral component in BHB that can be identified with the soft excess  in AGN. Moreover, the spectral transition in BHB takes place on viscous time scale, while the changing-look phenomena occurs on much shorter timescale compared to the viscous timescale \citep{2018MNRAS.480.3898N,2019MNRAS.483L..88P}). These dissimilarities suggest that changing-look phenomena of AGN and the spectral transition in BHB may not be caused by the same physical process.

\cite{2011MNRAS.414.2186J} have investigated different types of instabilities in accretion disks and argue that radiation pressure instability is likely to operate in AGN above an Eddington ratio of 0.025.  \cite{2019MNRAS.483L..88P} discussed the 2018 outburst of NGC~1566 in terms of the disk instability where the dominating radiation pressure over the gas pressure enhance the flux coming from the inner disk. \cite{2020A&A...641A.167S} explored the four bursts in NGC~1566 observed by \cite{1986ApJ...308...23A} during 1970--1985 with the disk instability mechanism, such that the disk becomes unstable during such events at the transition radius between the outer standard disk and the inner advection dominated accretion flow (ADAF) region. The rise time of the burst is expected to be comparable to the thermal timescale of the disk at the transition radius \citep{2020A&A...641A.167S}. However, a longer timescale of $\sim 9$ months has been claimed for the 2018 outburst than a typical thermal timescale of $\sim 1.7$ days at the inner disk radius of $50 R_g$. We found that the outburst timescale is comparable to the sound crossing time $\sim 1{\rm~year}$ at $\sim 50R_g$ assuming  gas pressure only. In case of the radiation pressure dominating over the gas pressure in the inner regions, the sound crossing time is expected to be even shorter but most likely  still  comparable to the outburst timescale. \citet{2011MNRAS.414.2186J} have studied the parameters space for the accretion disk instabilities, they found that the radiation pressure instability to be present in the inner region of AGN disks below $\sim 100R_g$ when  the Eddington ratio is larger than $0.025$. Interestingly, NGC~1566 seems to satisfy both these criteria as the soft excess emission observed during the outburst arises below $\sim 50R_g$ and Eddington fraction increased from $\sim 0.001$ (pre-outburst)  to $\sim 0.05$ (outburst peak) and then declined to $\sim 0.003$ when no soft excess is observed. Thus, we conclude that the 2018 outburst of NGC~1566 is most likely caused by the radiation pressure instability in the inner accretion disk. { This conclusion is valid for the adopted source distance of 21.3\mpc{} based on the Hubble parameter of $H_{0}= 70.3\km{}\s^{-1}\mpc^{-1}$ as well as for the source distance $16.9^{+7.5}_{-4.1}{\rm~Mpc}$ estimated from the Tully-Fisher relation \citep{2019MNRAS.487.2797E}. However, for the lower limit of the Tully-Fisher distance ($12.8\mpc{}$), the Eddington ratio ($L/L_{Edd} = 0.0126\pm0.0003$) for the June 2018 data is somewhat lower than the threshold Eddington rate for the disk instability. The Eddington ratio will be even much lower ($\sim 15$ times) for the shortest estimated distance of $5.5\mpc{}$ \citep{2014MNRAS.444..527S}, and in this case the disk instability condition can not be fulfilled. However, this distance is unlikely as it is excluded by the distance measurement based on the Tully-Fisher relation \citep{2019MNRAS.487.2797E}.} { Further, our conclusion is also based on the  adopted values of the black-hole spin parameter ($a = 0$)  and the  mass ($\log{M_{BH}/M_{\odot}} = 6.92$) for NGC~1566. We investigated the effect of the uncertainties in the estimation of black hole mass and spin parameter. We re-fitted the broadband UV/X-ray data of June 2018 using the lowest ($\log{M_{BH}/M_{\odot}} = 6.48$; \citealt{2015A&A...583A.104S}) and highest  ($\log{M_{BH}/M_{\odot}} = 7.11$;  \citealt{2014ApJ...789..124D}) estimated  masses and two different values of spin parameter ($a=0$ and $0.998$). 
%
The best-fit Eddington ratios for the largest black-hole mass are $L/L_{Edd} \sim 2.4\%$ for $a=0$, and $\sim 3.5\%$ for $a=0.998$. Similarly, the Eddington ratios for the smallest black-hole mass are $\sim 17\%$ for $a=0$, and $\sim 15\%$ for $a=0.998$. Thus, for any combinations of the mass and spin our conclusion remains valid, however for the largest black-hole mass and the lowest spin the Eddington ratio is just below but very close to the threshold value for the disk instability. }


\section{Conclusion}
\label{sec_conclusion}
{We characterized evolution of the primary spectral components -- the accretion disk, soft X-ray excess and the X-ray power law during the 2018 outburst of NGC~1566 by analyzing two FUV/X-ray data acquired by \astrosat{} during the decline phase of the outburst, and two UV/X-ray data acquired by \xmm{} before and at the peak of the outburst. We found that the accretion disk, soft X-ray excess, and X-ray power-law exhibited large amplitude variability during the outburst.  The variability of the soft X-ray excess component is significantly larger than those of the disk and the X-ray power-law components. Thus, the variability of the soft X-ray excess cannot be produced either by thermal Comptonization of the disk photons in a steady warm corona or by the disk illuminating X-ray power-law component, alone.  Rather, it is intrinsic to a changing warm corona. We refer this change in the warm corona as causing the spectral transition in NGC~1566 from a strong soft excess state to a negligible soft excess state. The outburst decline timescale is very different from the dynamical, thermal or viscous timescales of the accretion disk at the inner edge. The outburst  timescale is comparable to the sound crossing timescale of the disk at the transition radius of $\sim 50 R_g$ between the standard disk and warm Comptonizing region. We suggest that the transition of the source from a negligible soft X-ray excess state into the maximum soft X-ray excess state during the outburst is most likely caused by the radiation pressure instability in the inner regions of the disk. 

The declining X-ray power-law flux with the UV and the soft X-ray excess fluxes is due to thermal Comptonization of the disk and  soft excess photons in the hot corona. However, the X-ray power-law  does not steepen with increasing seed flux  which suggests intrinsic changes in the hot corona during the outburst. Thus, all three regions -- the accretion disk, the warm corona and the hot corona responsible for the primary continuum changed during the outburst. }

\acknowledgments

We thank an anonymous referee for constructive suggestions and comments. This publication uses the data from the AstroSat mission of the Indian Space Research Organisation (ISRO), archived at the Indian Space Science Data Centre (ISSDC). This research has made use of SXT data processing software provided by Tata Institute of Fundamental Research (TIFR), Mumbai, India. This research has made use of UVIT pipeline (CCDLAB). The UVIT data were checked and verified by the UVIT POC at IIA, Bnagalore, India. This research has made use of archival data of XMM–Newton provided by the XMM-Newton Science Archive developed by the ESAC Science Data Centre (ESDC) with requirements provided by the XMM-Newton Science Operations Centre. This research has made use of the {\it Python} and {\it Julia} packages. This research has made use of the NASA/IPAC Extragalactic Database (NED), which is operated by the Jet Propulsion Laboratory, California Institute of Technology, under contract with the National Aeronautics and Space Administration (NASA). 
 
\facilities{AstroSat, XMM-Newton}

\software{CCDLAB \citep{2017PASP..129k5002P},
	XSPEC \citep{1996ASPC..101...17A},
	Sherpa \citep{2001SPIE.4477...76F},
	SAOImageDS9 \citep{2003ASPC..295..489J},
Julia \citep{doi:10.1137/141000671},
	Astropy \citep{2013A&A...558A..33A}
		APLpy \citep{aplpy2012,aplpy2019}
}


\bibliography{mybib}
\bibliographystyle{aasjournal}

\end{document}